\documentclass[a4paper,11pt]{article}
\usepackage[utf8]{inputenc}
\usepackage{amsmath}
\usepackage{amsfonts}
\usepackage{amssymb}
\usepackage{comment}
\usepackage{graphicx}
\usepackage{tcolorbox}
\usepackage[toc]{appendix}
\usepackage{pdflscape}
\usepackage{color}
\usepackage{multirow}
\usepackage[normalem]{ulem}
\usepackage[left=2.00cm, right=2.00cm, top=2.00cm, bottom=2.00cm]{geometry}
\usepackage[colorlinks=true,citecolor=blue,urlcolor=cyan]{hyperref}
\allowdisplaybreaks
\usepackage{authblk}
\RequirePackage[numbers,sort&compress]{natbib}

\usepackage{mmacells}

\usepackage{eso-pic}
\usepackage{lipsum}

\newcommand{\g}[1]{\Gamma(#1)}

\newcommand{\eeqref}[1]{Eq. \eqref{#1}}
\newcommand{\mt}{\textsc{Mathematica}\,}

\newcommand{\fone}{\texttt{AppellF1.wl}\,}
\newcommand{\ftwo}{\texttt{AppellF2.wl}\,}
\newcommand{\fthree}{\texttt{AppellF3.wl}\,}
\newcommand{\lfs}{\texttt{LauricellaSaranFS.wl}\,}
\newcommand{\lfd}{\texttt{LauricellaFD.wl}\,}

\begin{document}
\title{Analytic continuations and numerical evaluation of the Appell $F_1$, $F_3$, Lauricella $F_D^{(3)}$ and Lauricella-Saran $F_S^{(3)}$ and their application to Feynman integrals}

\author[1,2]{Souvik Bera \thanks{\href{mailto:souvik.bera@apctp.org}{souvik.bera@apctp.org}}}
\author[1,3]{Tanay Pathak \thanks{\href{mailto:pathak.tanay@yukawa.kyoto-u.ac.jp}{pathak.tanay@yukawa.kyoto-u.ac.jp}}}
\affil[1]{Centre for High Energy Physics, Indian Institute of Science\\ Bangalore-560012, Karnataka, India.}
\affil[2]{Asia Pacific Center for Theoretical Physics,
77 Cheongam-ro, Nam-gu, Pohang-si, Gyeongsangbuk-do, 37673, Korea.}
\affil[3]{Center for Gravitational Physics and Quantum Information, Yukawa Institute for Theoretical Physics\\ Kyoto University, Kitashirakawa Oiwakecho, Sakyo-ku, Kyoto 606-8502, Japan.}
\date{}

\AddToShipoutPictureBG*{%
  \AtPageUpperLeft{%
  \vspace{5cm}
    \hspace{.85\paperwidth}%
    \raisebox{-3 \baselineskip}{%
      \makebox[10pt][r]{\\ YITP-24-85}
}}}%

\maketitle
\begin{abstract}
We present our investigation of the study of two variable hypergeometric series, namely Appell $F_{1}$ and $F_{3}$ series, and obtain a comprehensive list of its analytic continuations enough to cover the whole real $(x,y)$ plane, except on their singular loci. 
We also derive analytic continuations of their 3-variable generalization, the Lauricella $F_{D}^{(3)}$ series and the Lauricella-Saran $F_{S}^{(3)}$ series, leveraging the analytic continuations of $F_{1}$ and $F_{3}$, which ensures that the whole real $(x,y,z)$ space is covered, except on the singular loci of these functions. While these studies are motivated by the frequent occurrence of these multivariable hypergeometric functions in Feynman integral evaluation, they can also be used whenever they appear in other branches of mathematical physics. To facilitate their practical use, for analytical and numerical purposes, we provide four packages: $\texttt{AppellF1.wl}$, $\texttt{AppellF3.wl}$, $\texttt{LauricellaFD.wl}$, and $\texttt{LauricellaSaranFS.wl}$ in \textsc{Mathematica}.
These packages are applicable for generic as well as 
non-generic values of parameters, keeping in mind their utilities in the evaluation of the Feynman integrals. 
We explicitly present various physical applications of these packages in the context of Feynman integral evaluation and compare the results using other packages such as \texttt{FIESTA}. Upon applying the appropriate conventions for numerical evaluation, we find 
that the results obtained from our packages are consistent. Various \textsc{Mathematica} notebooks demonstrating different numerical results are also provided along with this paper.\\

\end{abstract}

\newpage

\setlength{\parindent}{20pt}
{\bf Summary of Programs:}
\begin{itemize} 
    \item \textit{Title of Programs}: \fone, \fthree, \lfd, \lfs.
    \vspace{-0.2cm}
    \item \textit{Licensing provisions}: GNU General Public License v.3.0.
    \vspace{-0.2cm}
    \item \textit{Programming language}: Wolfram Mathematica
    \item \textit{Repository }: \url{https://github.com/souvik5151/Appell_Lauricella_Saran_functions}.
    \item \textit{Nature of problem}: To find and develop a numerically consistent implementation of the analytic continuations of various two and three-variable hypergeometric functions, namely Appell $F_{1}$, $F_{3}$, Lauricella $F_{D}^{(3)}$ and Lauricella-Saran $F_{S}^{(3)}$ that typically appear in the evaluation of Feynman integrals. 
    
 \vspace{-0.2cm}
    \item \textit{Solution method}: Use the method of Olsson to find the analytic continuation of these functions and to implement these analytic continuations, following appropriate convention, in \mt for consistent numerical evaluation. For the values of the Pochhammer parameters corresponding to the non-generic cases, a proper limiting procedure is implemented internally. 
\end{itemize}
\newpage

\section{Introduction}


In this paper, we present our study on the analytic continuations of multivariable hypergeometric functions. In particular we study the two-variable Appell $F_1$ and $F_3$ \cite{Appell1880}  and three-variable Lauricella-Saran $F_D^{(3)}$ and $F_S^{(3)}$  \cite{LauricellaFs,Saran_transhypgeo}  series. These investigations are a continuation of various other works carried out for various multivariable hypergeometric series such as Appell $F_{1}$ \cite{Olsson64,Bezrodnykh2017AnalyticCO}, Appell $F_{2}$ \cite{Olsson77,Ananthanarayan:2021bqz}, Appell $F_{4}$ \cite{Exton_1995, Huber:2007zz, Ananthanarayan:F4}, Horn $H_{1}$ and $H_{5}$ \cite{Bera:2022eag}, Kamp\'e de F\'eriet series \cite{bezrodnykh2022analytic3} and Lauricella series $F_{D}^{(N)}$ \cite{Exton:1976,bezrodnykh2016analytic, Bezrodnykh:LauricellaFD,bezrodnykh2022analytic}, Srivastava $H_C^{(3)}$ \cite{Friot:2022dme} and  other multivariable Horn functions \cite{bezrodnykh2022formulas,bezrodnykh2022analytic3, bezrodnykh2020analytic}. Standard references of various properties and expressions related to multivariate hypergeometric functions can be found in \cite{Slater:1966, bateman,Exton:1976,Srivastava:1985,bailey1935generalized,schlosser2013multiple}.

The motivation behind studying these particular functions comes from their ubiquitous presence in physics and mathematical physics, notably within the domain of Feynman integral calculus. In particular, it has been shown that almost all the Feynman integral can be written in terms of a general system of a hypergeometric function called the GKZ /$\mathcal{A}$- hypergeometric system \cite{gel1992general,newstead1995discriminants,gel1987holonomic,gelfand1991hypergeometric,gelfand1990generalized, delaCruz:2019skx,Klausen:2019hrg}, which is now a very active field of research \cite{Feng:2019bdx,Klausen:2021yrt,Feng:2022kgh,Agostini:2022cgv,Chestnov:2022alh,Feng:2022ude,Zhang:2023fil,Chestnov:2023kww,Matsubara-Heo:2023ylc}.  There also exists computer program that can provide hypergeometric function representation of a given Feynman integral \cite{Ananthanarayan:2022ntm}.

The hypergeometric functions have a long history of appearing in the evaluation of Feynman integrals. using Mellin-Barnes representations \cite{Dubovyk:2022obc}, in \cite{Davydychev:1990cq,Davydychev:1990jt},  some general results for massive and massless $N-$point functions are obtained in terms of multivariable hypergeometric functions, including $N-$ variable Lauricella $F_{D}^{(N)}$. Moreover, the general massive one-loop Feynman integral can be represented as a meromorphic function of space-time dimensions using Gauss hypergeometric function $_{2}F_{1}$, Appell $F_{1}$ and Lauricella-Saran function $F_{S}^{(3)}$ for self-energy, vertex and box integrals, respectively \cite{Tarasov:2000sf,Fleischer:2003rm,Phan:2018cnz,Riemann:2019gqf}. For the one loop pentagon integral in the Regge limit, the result is expressed in terms of Appell $F_{4}$ and some Kamp\'e de F\'eriet functions \cite{DelDuca:2009ac}. At the two loop level, Appell $F_{4}$ and its 3-variable generalization Lauricella $F_{C}^{(3)}$ appear in the study of sunset integrals \cite{Berends:1993ee,Ananthanarayan:2019icl}. Various other application and occurrence of these functions can be found in \cite{Tarasov:2006nk,Feng:2018zxf,Gu:2018aya,Gu:2020ypr,Feng:hypergeometry,Duhr:2023bku}. In fact,
the relation between hypergeometric functions and Feynman integrals is bidirectional. A number of new relations are discovered while studying the hypergeometric functions representation of Feynman integrals \cite{Fleischer:1999hq, Tarasov:2015wcd, Kniehl:2011ym,Tarasov:2022clb,Blumlein:2021hbq, Ananthanarayan:2023nol,Shpot:2007bz}. The application of these functions is also widespread in other fields of physics, such as in conformal field theory \cite{Fan:2023lky,Akerblom:2004cg,isachenkov2016conformal}, where they occur quite frequently.

We focus on the evaluation of these multivariable hypergeometric functions outside their defining domain of convergence. In the hypergeometric function representation, the ratio of the scales involved in a Feynman integral appears in the variables of the hypergeometric functions. Thus, these analytic continuations help to evaluate these Feynman integrals for any kinematical values of choice. To derive the analytic continuations of Appell $F_1$ and $F_3$ and Lauricella-Saran $F_D^{(3)}$ and $F_S^{(3)}$, we use the method of Olsson  \cite{Olsson64}. This approach leverages the transformation properties of hypergeometric functions with fewer variables to determine the analytic continuations of the specified hypergeometric function. The method has been extensively used to study and obtain the analytic continuations of two variable Appell $F_{1}$ \cite{Olsson64} and its $N-$variable generalization, i.e., Lauricella $F_{D}^{(N)}$ \cite{Exton:1976}, Appell $F_{4}$ \cite{Exton_1995, Huber:2007zz}, Appell $F_{2}$ \cite{Ananthanarayan:2021bqz} and Horn  $H_{1}$ and $H_{5}$ \cite{Bera:2022eag}. The analytic continuations of Appell $F_{1}$ have been implemented for numerical purposes in \texttt{Fortran}, which is called \texttt{f1} \cite{Colavecchia2001,Colavecchia2004}. While Appell $F_{2}$ is also implemented as the \ftwo \cite{Ananthanarayan:2021bqz} package in \mt. The method of Olsson itself has also been automated in the form of a package named \texttt{Olsson.wl} \cite{Ananthanarayan:2021yar} in \mt, which we use heavily in our present study. The package makes, the otherwise laborious and error prone task of deriving analytic continuations easy, quick and error-free.

Our study aims to achieve consistent and efficient evaluation of the derived analytic continuations. Consistency refers to the property that different analytic continuations that converge at the same point yield the same result. This is theoretically expected, but in practice, due to multi-valuedness, different analytic continuations may be evaluated on different sheets and produce different results. We address this problem by following the approach used for \texttt{AppellF2.wl} in \cite{Ananthanarayan:2021bqz}. Another important aspect of the packages is the evaluation of the analytic continuations for the non-generic values of Pochhammer parameters. Previously, in the \texttt{AppellF2.wl} package, this issue was handled manually via a proper limiting procedure. While in all the packages that we present in this paper the necessary steps are incorporated internally, which makes them applicable for non-generic values of Pochhammer parameters as well. For the efficient evaluations of the finite series summation, we use techniques to reduce the number of operations needed to perform the finite summations, as well as internal commands of \mt that have a fast and efficient implementation. These techniques effectively reduce the number of operations through certain transformations of summation indices that we discuss in later sections. We observe significant improvement in time and accuracy in the evaluation and results obtained. Using these techniques, the analytic continuations form the core of the widely used computer program, \mt based packages \fone, \fthree, \lfd and \lfs, which can numerically evaluate these functions for any values of Pochhammer parameters and real values of its arguments, except the singular loci of associated functions. Since all the analytic continuations of all the functions are stored inside the packages, this also enables the use of the packages for analytical purposes as well.

The article is organized as follows: The definitions and useful relations of Appell $F_1, F_3$, Lauricella  $F_D^{(3)}$ and Lauricella-Saran $F_S^{(3)}$ are provided in Section \ref{sec:definitions}. The method of Olsson for deriving the analytic continuations of multivariate hypergeometric functions is reviewed with an example of Lauricella $F_S^{(3)}$, in Section \ref{sec:methodolsson}. The algorithm of these packages is discussed in Section \ref{sec:algorithm}, which is followed by a demonstration of the  commands of the packages in Section \ref{sec:demo}. The numerical tests of these packages are presented in Section \ref{sec:numtest}. The tests of these packages for non-generic values of the parameters, motivated by the examples of Feynman integrals, are detailed in Section \ref{sec:application}. The summary and conclusions are drawn in Section \ref{sec:summary}. All the packages and several test files can be found in the \href{https://github.com/souvik5151/Appell_Lauricella_Saran_functions}{GitHub} repository. A list of definitions of relevant hypergeometric series and a section on error estimation are provided in appendices \ref{appendix:def} and \ref{appendix:error} respectively. 

\section{Definitions}
\label{sec:definitions}
In this Section, we provide the definitions and useful expressions related to the functions of our study. We start with double variable Appell functions.

\subsection{Appell Functions: $F_{1}$ and $F_{3}$}


The four Appell functions \cite{Appell1880} are the two-variable generalizations of the Gauss hypergeometric functions $_2F_1(a,b;c|x)$, whose series representation is given by
\begin{align}
    _2F_1(a,b;c|x) = \sum_{m= 0 }^\infty \frac{(a)_m (b)_m}{(c)_m} \frac{x^m}{m!}, \hspace{1cm} |x|<1
\end{align}
where $(a)_m = \frac{\Gamma(a+m)}{\Gamma(a)}$ is the Pochhammer symbol.

The analytic continuations of Gauss $_2F_1(\dots| x)$ around $x=1$ and $x=\infty$ are well-known \cite{Srivastava:1985,bateman}

\begin{align}
		{ }_{2} F_{1}(a, b, c | x) &=\frac{\Gamma(c) \Gamma(c-a-b)}{\Gamma(c-a) \Gamma(c-b)}{ }_{2} F_{1}(a, b, a+b-c+1 | 1-x) \nonumber\\
		&+\frac{\Gamma(c) \Gamma(a+b-c)}{\Gamma(a) \Gamma(b)}(1-z)^{c-a-b}{ }_{2} F_{1}(c-a, c-b, c-a-b+1 | 1-x) \label{ac_2f1at1}
\end{align}
which is valid for $|1-x| < 1$ and

\begin{align}
		{ }_{2} F_{1}(a, b, c | x) &=\frac{\Gamma(c) \Gamma(b-a)}{\Gamma(b) \Gamma(c-a)}(-z)^{-a}{ }_{2} F_{1}\left(a, a-c+1, a-b+1 |\frac{1}{x}\right) \nonumber\\
		&+\frac{\Gamma(c) \Gamma(a-b)}{\Gamma(a) \Gamma(c-b)}(-z)^{-b}{ }_{2} F_{1}\left(b, b-c+1, b-a+1 | \frac{1}{x}\right)  \label{ac_2f1atinf}
\end{align}
which is valid for $\left|\frac{1}{x}\right| < 1$.

The four Appell functions are defined by the following series representation

\begin{align}\label{eqn:appellfun}
    F_1\left(a, b_1, b_2 ; c | x, y\right) &=\sum_{m, n=0}^{\infty} \frac{(a)_{m+n}\left(b_1\right)_m\left(b_2\right)_n}{(c)_{m+n} }  \frac{x^m}{m!}\frac{y^n}{n!},\hspace{1cm} |x|<1\wedge|y|<1 \\ 
    F_2\left(a, b_1, b_2 ; c_1, c_2 | x, y\right)&=\sum_{m, n=0}^{\infty} \frac{(a)_{m+n}\left(b_1\right)_m\left(b_2\right)_n}{\left(c_1\right)_m\left(c_2\right)_n } \frac{x^m}{m!}\frac{y^n}{n!},\hspace{1cm}  |x|+|y|<1 \\
    F_3\left(a_1, a_2, b_1, b_2 ; c | x, y\right)&=\sum_{m, n=0}^{\infty} \frac{\left(a_1\right)_m\left(a_2\right)_n\left(b_1\right)_m\left(b_2\right)_n}{(c)_{m+n} } \frac{x^m}{m!}\frac{y^n}{n!},\hspace{1cm} |x|<1\wedge|y|<1 \\
    F_4\left(a, b ; c_1, c_2 | x, y\right)&=\sum_{m, n=0}^{\infty} \frac{(a)_{m+n}(b)_{m+n}}{\left(c_1\right)_m\left(c_2\right)_n } \frac{x^m}{m!}\frac{y^n}{n!},\hspace{1cm} \sqrt{|x|}+ \sqrt{|y|}<1
\end{align}

The analysis of three out of the four Appell functions has been done previously. The analytic continuations of Appell $F_{1}$ are obtained in \cite{Olsson64}, while in \cite{Exton_1995, Huber:2007zz},  Appell $F_{4}$ is analysed in a similar manner and   Appell $F_{2}$ is treated in \cite{Ananthanarayan:2021bqz}. A package \ftwo for the numerical implementation of $F_{2}$ in \textsc{Mathematica} is also presented in \cite{Ananthanarayan:2021bqz}, along with the usual analysis and evaluation of analytic continuations. There already exist a numerical program to evaluate Appell $F_1$  \cite{Colavecchia2001, Colavecchia2004}. It should also be noted that all the Appell functions have been implemented in \textsc{Maple} \cite{maple2022} and in \mt v.13.3 \cite{Mathematica13.3} recently. However, the numerical tests performed and presented in Section \ref{sec:numtest}, show that the numerical evaluation of these functions using these implementations is still far from perfect. Therefore, the present analysis is timely and may be useful for improving such implementations further.

Apart from the above series representations, the Appell functions also have Euler integral representations that can be used for analytic continuation purposes. Since our interest lies in the analytic continuations of Appell $F_{1}$ and $F_{3}$, we only present the integral representations of these, which are as follows. 

\begin{equation}
F_1\left(a, b_1, b_2, c | x, y\right)=\frac{\Gamma(c)}{\Gamma(a) \Gamma(c-a)} \int_0^1 t^{a-1}(1-t)^{c-a-1}(1-x t)^{-b_1}(1-y t)^{-b_2} \mathrm{~d} t, \quad \Re (c)>\Re (a)>0 .
\end{equation}
and,
\begin{align}
 F_3\left(a_{1}, a_{2}, b_{1}, b_{2} ; c | x, y\right) &= 
\frac{\Gamma(c)}{\Gamma(b_{1}) \Gamma\left(b_{2}\right) \Gamma\left(c-b_{1}-b_{2}\right)} \nonumber\\
&\times \iint u^{b-1} v^{b_{2}-1}(1-u-v)^{c-b_{1}-\beta^{\prime}-1}(1-u x)^{-a_{1}}(1-v y)^{-a_{2}} \mathrm{d} u \mathrm{~d} v,
\end{align}
where integration domain is given by: $u \geq 0, \quad v \geq 0, \quad u+v \leq 1$, \\
and the integral is convergent for:
$\, \Re(b_{1})>0,\,
\Re\left(b_{2}\right)>0, \, \Re\left(c-b_{1}-b_{2}\right)>0$.

Another important property of interest to us is the knowledge of the singular loci of these hypergeometric functions, where we omit the evaluation of these functions in general. 
For Appell $F_{1}$ the singular loci are found to be \cite{Olsson64}
\begin{align}
    x=0\lor y=0\lor x=1\lor y=1\lor x=y
\end{align}

Similarly for Appell $F_{3}$ they are as follows \cite{erdelyi1951analytic} 
\begin{align}
    x=0\lor y=0\lor x=1\lor y=1\lor x y = x+y 
\end{align}



\subsection{Lauricella Function: $F_{D}^{(N)}$}
Lauricella further generalized the four Appell functions to $N-$variables in \cite{LauricellaFs}. 
In this Section, we provide the definition and some basic properties of one such function of interest to us, i.e., Lauricella $F_{D}^{(N)}$ which is the generalization of double variable Appell $F_1$. The Lauricella  $F_{D}^{(N)}$, in $N-$variables, is defined by the following series representation
\begin{equation}\label{eq:lfd}
    F_{D}^{(N)}(a,b_{1}, \cdots, b_{N};c|z_{1},\cdots,z_{N}) = \sum_{j_{1}=0}^{\infty} \cdots \sum_{j_{N}=0}^{\infty} \frac{(a)_{j_{1}+\cdots + j_{N}}(b_{1})_{j_1} \cdots (b_N)_{j_N} }{(c)_{j_{1}+ \cdots + j_{N}}} \frac{z_{1}^{j_{1}} \cdots z_{N}^{j_{N}}}{j_{1}! \cdots j_{N}!}
\end{equation}
which is valid for $|z_{1}| < 1 \wedge \cdots \wedge |z_{N}| <1$.

The function remarkably has a one-dimensional Euler-type integral representation given by
\begin{align}
    F_D^{(N)}\left(a, b_1, \ldots, b_n, c | x_1, \ldots, x_n\right)=\frac{\Gamma(c)}{\Gamma(a) \Gamma(c-a)} \int_0^1 t^{a-1}(1-t)^{c-a-1}\left(1-x_1 t\right)^{-b_1} \cdots\left(1-x_n t\right)^{-b_n} \mathrm{~d} t, .
\end{align}
which is convergent for  $\Re(c) > \Re (a) > 0$.


For $N=2$, Eq. \eqref{eq:lfd} reduces to Appell $F_{1}$ series
\begin{equation}
F_1(a,b_{1},b_{2};c|z_{1},z_{2}) = F_{D}^{(2)}(a,b_{1},b_{2};c|z_{1},z_{2}) = \sum_{j_{1}=0}^{\infty} \sum_{j_{2}=0}^{\infty} \frac{(a)_{j_{1}+ j_{2}}(b_{1})_{j_1}(b_2)_{j_2} }{(c)_{j_{1}+ j_{2}}} \frac{z_{1}^{j_{2}} z_{2}^{j_{2}}}{j_{2}! j_{2}!}
\end{equation}

For $N=3$, Eq. \eqref{eq:lfd} reduces to Lauricella $F_{D}^{(3)}$ series
\begin{equation}\label{eq:lfd3}
    F_{D}^{(3)}(a,b_{1},b_{2}, b_3;c|z_{1},z_{2},z_{3}) = \sum_{j_{1}=0}^{\infty} \sum_{j_{2}=0}^{\infty}\sum_{j_{3}=0}^{\infty} \frac{(a)_{j_{1}+ j_{2}+j_{3}}(b_{1})_{j_1}(b_2)_{j_2} (b_{3})_{j_{3}}}{(c)_{j_{1}+ j_{2}+j_{3}}} \frac{z_{1}^{j_{2}} z_{2}^{j_{2}} z_{3}^{j_{3}}}{j_{2}! j_{2}! j_{3}!}
\end{equation}

The singular loci of $F_{D}^{(3)}$ are given by,
\begin{align}
    x=0\lor y=0\lor z=0\lor  x=1\lor y=1\lor z=1 \lor x=y \lor y = z  \lor x = z
\end{align}

\subsection{Lauricella-Saran function : $F_{S}^{(3)}$} \label{sec:lsaran}
Another three-variable hypergeometric series of interest to us is the Lauricella-Saran $F_{S}^{(3)}$ \cite{Saran_transhypgeo}. It is defined as


\begin{align}\label{eqn:LS_def}
 F_S^{(3)} :=  F_S^{(3)} (a_1,a_2,b_1,b_2,b_3;c|x,y,z) = \sum_{m,n,p=0}^\infty \frac{ \left(a_1\right)_m \left(a_2\right)_{n+p} \left(b_1\right)_m \left(b_2\right)_n \left(b_3\right)_p }{ (c)_{m+n+p}} \frac{x^m y^n z^p}{m! n! p!}
\end{align}
The defining series representation of the $F_{S}^{(3)}$ is valid in \cite{Srivastava:1985}
\begin{align}
    |x| < 1 \wedge |y|<1 \wedge |z|<1
\end{align}

The Lauricella-Saran function $F_{S}^{(3)}$ also has a triple Euler type integral representation as follows
\begin{align} 
F_S^{(3)}  &= \frac{\Gamma\left(c\right)}{\Gamma\left(b_1\right)\Gamma\left(b_2\right) \Gamma\left(b_3\right) \Gamma\left(c-b_1-b_2-\beta_3\right)}\nonumber\\
&\times \iiint u^{b_1-1} v^{b_2-1} w^{b_3-1} 
  (1-u-v-w)^{c-b_1-b_2-b_2-1} (1-u x)^{-a_1}(1-v y-w z)^{-a_2} du\, dv\, dw .
\end{align}

which is convergent for  $\Re(c) > \Re (b_{1}+b_{2}+b_{3}) > 0$, $\Re(c) > \Re (b_{1}) > 0$, $\Re (b_{2}) > 0$, $\Re (b_{3}) > 0$ .

The singular loci of the Lauricella-Saran function are found to be \cite{Hyperdire2}
\begin{align}
    x=0\lor y=0\lor z=0\lor x=1\lor y=1\lor z=1\lor y=z\lor x+y=x y\lor x+z=x z
\end{align}

We would like to remark that we only list some basic properties of all these functions of interest. For other properties of these functions, such as the system of the partial differential equation they satisfy, symmetry relations, transformation properties, etc, we refer the readers to the standard references \cite{Srivastava:1985,bateman,Slater:1966, bailey1935generalized,Exton:1976,LauricellaFs,erdelyi_1948,schlosser2013multiple}.

\section{The Method of Olsson} \label{sec:methodolsson}
Appell and Kamp\'e de Feriet \cite{appellpandkampe1926fonctions} have
 pointed out that analytic continuations of a multivariable hypergeometric function can be derived by leveraging the known analytic continuations of the hypergeometric functions with variables lower in number than that of the former function. In \cite{Olsson64}, Olsson has obtained the analytic continuations of the double variable Appell $F_1$ function \cite{Appell1880} employing the linear transformation formulae of the Gauss $_2F_1$ function. The same technique has been used to find the analytic continuations of the Appell $F_2$ recently, which are the backbone of the numerical package \texttt{AppellF2.wl} \cite{Ananthanarayan:2021bqz}. This powerful technique to find the analytic continuations of multivariable hypergeometric functions is automated in the \mt based package \texttt{Olsson.wl} \cite{Ananthanarayan:2021yar}, which eases the calculations significantly.

In this Section, to illustrate the method, we find the analytic continuations of the Lauricella-Saran function around $(0,\infty,\infty)$ and $(\infty,\infty,\infty)$ using the method of Olsson. 

One can immediately observe the following symmetry of the $F_{S}$.
\begin{align}\label{eqn:LS_symmetry}
    F_S^{(3)} (a_1,a_2,b_1,b_2,b_3;c|x,y,z) = F_S^{(3)} (a_1,a_2,b_1,b_3,b_2;c|x,z,y)
\end{align}

We observe that summing over one of the summation indices, one can write the summand in terms of double variable Appell $F_1$ or $F_3$ function
\begin{align}
    F_S^{(3)} (a_1,a_2,b_1,b_2,b_3;c| x,y,z) &= \sum_{m=0}^\infty \frac{ \left(a_1\right)_m \left(b_1\right)_m }{ (c)_m} \frac{x^m}{m!} F_1\left(a_2;b_2,b_3;c+m | y,z\right) \label{eqn:LS_F1}\\
    &= \sum_{n=0}^\infty \frac{ \left(a_2\right)_n \left(b_2\right)_n }{ (c)_n} \frac{y^n}{n!}F_3\left(a_1,b_3,b_1,a_2+n;c+n|x,z\right) \\
    &= \sum_{p=0}^\infty \frac{ \left(a_2\right)_p \left(b_3\right)_p }{ (c)_p} \frac{z^p}{p!}F_3\left(a_1,b_2,b_1,a_2+p;c+p|x,y\right)
\end{align}

Thus, a good use of the analytic continuations of the Appell functions can be made to find the analytic continuations of $F_S^{(3)}$. To find the analytic continuation of the Lauricella-Saran function around $(0,\infty,\infty)$, we use the analytic continuation of the Appell $F_1$ around $(\infty,\infty)$ from \cite{Olsson64} (precisely  Eq. (22) of the \cite{Olsson64}), which is shown below.
 \begin{align}\label{eqn:F1_inf_inf}
 &F_1(a,b_1,b_2;c|x,y) \nonumber \\&= (-x)^{-b_1} (-y)^{b_1-a} \frac{ \Gamma (c)  \Gamma \left(a-b_1\right) \Gamma \left(-a+b_1+b_2\right) }{\Gamma (a) \Gamma \left(b_2\right) \Gamma (c-a)} ~ G_2 \left(b_1,a-c+1,a-b_1,-a+b_1+b_2 \Big| -\frac{y}{x},-\frac{1}{y}\right)\nonumber\\
    &+ (-x)^{-a} \frac{ \Gamma (c) \Gamma \left(b_1-a\right) }{\Gamma \left(b_1\right) \Gamma (c-a)} F_1\left(a,a-c+1,b_2;a-b_1+1 \Big|\frac{1}{x},\frac{y}{x}\right)\nonumber\\
    &+(-x)^{-b_1} (-y)^{-b_2} \frac{ \Gamma (c) \Gamma \left(a-b_1-b_2\right) }{\Gamma (a) \Gamma \left(c-b_1-b_2\right)} F_1\left(b_1+b_2-c+1,b_1,b_2;-a+b_1+b_2+1 \Big| \frac{1}{x},\frac{1}{y}\right)
 \end{align}
 where $G_2$ is the one of the Horn's functions (for definition see Appendix \ref{appendix:def}). The domain of convergence of the above analytic continuation is given by $\frac{1}{| x| }<1\land \frac{1}{| y| }<1\land \left| \frac{y}{x}\right| <1$.

 Inserting the above expression in \eeqref{eqn:LS_F1} and simplifying the gamma functions and Pochhammer symbols, we obtain the analytic continuation of $F_S^{(3)}$ around $(0,\infty,\infty)$

 \begin{align}\label{eqn:LS_0infinf}
     F_S^{(3)} &= (-y)^{-a_2} \frac{ \Gamma (c) \Gamma \left(b_2-a_2\right)}{\Gamma \left(b_2\right) \Gamma \left(c-a_2\right)}\sum_{m,n,p=0}^\infty  
     \frac{ \left(a_1\right)_m \left(b_1\right)_m \left(b_3\right)_p \left(a_2\right)_{n+p} \left(-c+a_2+1\right)_{n-m}}{ \left(a_2-b_2+1\right)_{n+p}} \frac{(-x)^m  y^{-n-p} z^p }{m! n! p!} \nonumber\\
     &+(-y)^{-b_2} (-z)^{b_2-a_2} \frac{ \Gamma (c)  \Gamma \left(a_2-b_2\right) \Gamma \left(-a_2+b_2+b_3\right)}{\Gamma \left(a_2\right) \Gamma \left(b_3\right) \Gamma \left(c-a_2\right)}\nonumber\\
     &\times \sum_{m,n,p=0}^\infty  \frac{\left(a_1\right)_m \left(b_1\right)_m \left(b_2\right)_n  \left(a_2-b_2\right)_{p-n} \left(-c+a_2+1\right)_{p-m}}{ \left(a_2-b_2-b_3+1\right)_{p-n}} \frac{(-x)^m  y^{-n} z^{n-p} }{m! n! p!}\nonumber\\
     &+ (-y)^{-b_2} (-z)^{-b_3} \frac{ \Gamma (c) \Gamma \left(a_2-b_2-b_3\right)}{\Gamma \left(a_2\right) \Gamma \left(c-b_2-b_3\right)} \nonumber\\
     &\times \sum_{m,n,p=0}^\infty \frac{ \left(a_1\right)_m \left(b_1\right)_m \left(b_2\right)_n \left(b_3\right)_p \left(-c+b_2+b_3+1\right)_{-m+n+p}}{ \left(-a_2+b_2+b_3+1\right)_{n+p}} \frac{(-x)^m y^{-n} z^{-p}}{m! n! p!}
 \end{align}
 which can be written in terms of other triple variable hypergeometric functions (see Appendix \ref{appendix:def} for definitions) as,

\begin{small}
 
 \begin{align*}
     F_S^{(3)}  &= (-y)^{-a_2} \frac{ \Gamma (c) \Gamma \left(b_2-a_2\right)}{\Gamma \left(b_2\right) \Gamma \left(c-a_2\right)} \times F_{5c}\left(a_2,b_3,a_1,b_1,1+a_2-c;1+a_2-b_2\left| \frac{z}{y},\frac{1}{y},-x\right.\right) \nonumber\\
    &+  (-y)^{-b_2}  (-z)^{b_2-a_2} \frac{ \Gamma (c) \Gamma \left(a_2-b_2\right) \Gamma \left(-a_2+b_2+b_3\right)}{\Gamma \left(a_2\right) \Gamma \left(b_3\right) \Gamma \left(c-a_2\right)} \nonumber\\
    &\times F_{1e}\left(b_2,a_1,b_1,b_2+b_3-a_2, 1+a_2-c,a_2-b_2\left|-\frac{z}{y},-\frac{1}{z},-x\right.\right) \nonumber\\
    &+ (-y)^{-b_2} (-z)^{-b_3}  \frac{ \Gamma (c) \Gamma \left(a_2-b_2-b_3\right)}{\Gamma \left(a_2\right) \Gamma \left(c-b_2-b_3\right)}  \times F_{5b}\left(b_3, b_2, b_1, a_1, 1+b_2+b_3-c; 1-a_2+b_2+b_3\left| \frac{1}{z},\frac{1}{y}, -x\right.\right)
 \end{align*}
   
\end{small}
 
 The domain of convergence can be found using the Horn's theorem \cite{Srivastava:1985}. The above analytic continuation is valid in:
 \begin{align}
     | x| <1\land \frac{1}{| y| }+1<\frac{1}{| x| }\land \frac{1}{| y| }<1\land \frac{1}{| z| }+1<\frac{1}{| x| }\land \frac{1}{| z| }<1\land \left| \frac{z}{y}\right| <1
 \end{align}

One can immediately find another analytic continuation of the $F_S^{(3)}$  around $(0,\infty,\infty)$ using the symmetry relation (i.e., \eeqref{eqn:LS_symmetry}), which we do not present here explicitly. 

Next, we find the analytic continuation around $(\infty,\infty,\infty)$ by taking each of the three series of \eeqref{eqn:LS_0infinf}, summing over the index $m$ and using the analytic continuation of $_2F_1(x)$ around $x=\infty$ (i.e., \eeqref{ac_2f1atinf}). Let us denote the first series of \eeqref{eqn:LS_0infinf} as $S_1$
\begin{align}
    S_1 &= \sum_{m,n,p=0}^\infty  
     \frac{ \left(a_1\right)_m \left(b_1\right)_m \left(b_3\right)_p \left(a_2\right)_{n+p} \left(-c+a_2+1\right)_{n-m}}{ \left(a_2-b_2+1\right)_{n+p}} \frac{(-x)^m  y^{-n-p} z^p }{m! n! p!}\nonumber\\
     &= \sum_{n,p=0}^\infty \frac{ \left(b_3\right)_p  \left(-c+a_2+1\right)_n \left(a_2\right)_{n+p} }{\left(a_2-b_2+1\right)_{n+p}} \frac{y^{-n-p} z^p}{n! p! } \, _2F_1\left(a_1,b_1;c-n-a_2|x\right)
\end{align}

In the second equality, the summation over the index $m$ is explicitly taken. As a result, the Gauss $_2F_1$ appears in the summand. 
Now we use \eeqref{ac_2f1atinf} to find the analytic continuation of the series $S_1$, which is labelled as $S_1'$ below
\begin{align}\label{eqn:LS_infinfinf1}
    S_1' &= (-x)^{-a_1} \frac{ \Gamma \left(b_1-a_1\right) \Gamma \left(c-a_2\right)}{\Gamma \left(b_1\right) \Gamma \left(c-a_1-a_2\right)}  \sum_{m,n,p=0}^\infty \frac{ \left(a_1\right)_m \left(b_3\right)_p  \left(a_2\right)_{n+p} \left(-c+a_1+a_2+1\right)_{m+n}}{\left(a_1-b_1+1\right)_m \left(a_2-b_2+1\right)_{n+p}} \frac{x^{-m} y^{-n-p} z^p}{m! n! p! }\nonumber\\
    &+ (-x)^{-b_1} \frac{ \Gamma \left(a_1-b_1\right) \Gamma \left(c-a_2\right)}{\Gamma \left(a_1\right) \Gamma \left(c-a_2-b_1\right)} \sum_{m,n,p=0}^\infty \frac{ \left(b_1\right)_m \left(b_3\right)_p  \left(a_2\right)_{n+p} \left(-c+a_2+b_1+1\right)_{m+n}}{ \left(-a_1+b_1+1\right)_m \left(a_2-b_2+1\right)_{n+p}} \frac{x^{-m} y^{-n-p}  z^p}{m! n! p!}\\
    &=(-x)^{-a_1} \frac{ \Gamma \left(b_1-a_1\right) \Gamma \left(c-a_2\right)}{\Gamma \left(b_1\right) \Gamma \left(c-a_1-a_2\right)} ~F_M \left(a_2, 1+a_1+a_2-c,b_3,a_1; 1+a_2-b_2,1+a_1-b_1\left| \frac{z}{y}, \frac{1}{y}, \frac{1}{x} \right.
    \right)  \nonumber\\
    &+(-x)^{-b_1} \frac{ \Gamma \left(a_1-b_1\right) \Gamma \left(c-a_2\right)}{\Gamma \left(a_1\right) \Gamma \left(c-a_2-b_1\right)}  ~F_M \left(a_2, 1+b_1+a_2-c,b_3,b_1; 1+a_2-b_2,1+b_1-a_1\left| \frac{z}{y}, \frac{1}{y}, \frac{1}{x} \right.
    \right)  \nonumber
\end{align}
Similarly, taking the second series of \eeqref{eqn:LS_0infinf} (which we denote by $S_2$) and proceeding as before
\begin{align}
    S_2 &= \sum_{m,n,p=0}^\infty \frac{ \left(a_1\right)_m \left(b_1\right)_m \left(b_2\right)_n  \left(a_2-b_2\right)_{p-n} \left(-c+a_2+1\right)_{p-m}}{\left(a_2-b_2-b_3+1\right)_{p-n}} \frac{(-x)^m y^{-n} z^{n-p}}{m! n! p! }\nonumber\\
    &= \sum_{n,p=0}^\infty \frac{ \left(b_2\right)_n  \left(-c+a_2+1\right)_p \left(a_2-b_2\right)_{p-n} }{ \left(a_2-b_2-b_3+1\right)_{p-n}} \frac{y^{-n} z^{n-p} }{n! p!}  \, _2F_1\left(a_1,b_1;c-p-a_2|x\right)
\end{align}
Again using the analytic continuation of $_2F_1 (\dots ; x)$ around $x=\infty$, we obtain the analytic continuation of $S_2$ as
\begin{small}

\begin{align}\label{eqn:LS_infinfinf2}
    S_2' &=(-x)^{-a_1} \frac{ \Gamma \left(b_1-a_1\right) \Gamma \left(c-a_2\right)}{\Gamma \left(b_1\right) \Gamma \left(c-a_1-a_2\right)} \sum_{m,n,p=0}^\infty \frac{ \left(a_1\right)_m \left(b_2\right)_n  \left(a_2-b_2\right)_{p-n} \left(-c+a_1+a_2+1\right)_{m+p}}{ \left(a_1-b_1+1\right)_m \left(a_2-b_2-b_3+1\right)_{p-n}} \frac{x^{-m} y^{-n} z^{n-p}}{m! n! p!}\nonumber\\
    &+ (-x)^{-b_1} \frac{ \Gamma \left(a_1-b_1\right) \Gamma \left(c-a_2\right)}{\Gamma \left(a_1\right) \Gamma \left(c-a_2-b_1\right)}  \sum_{m,n,p=0}^\infty
    \frac{ \left(b_1\right)_m \left(b_2\right)_n  \left(a_2-b_2\right)_{p-n} \left(-c+a_2+b_1+1\right)_{m+p}}{ \left(-a_1+b_1+1\right)_m \left(a_2-b_2-b_3+1\right)_{p-n}} \frac{x^{-m} y^{-n} z^{n-p}}{m! n! p!}
    \\
    \nonumber\\
    &=(-x)^{-a_1} \frac{ \Gamma \left(b_1-a_1\right) \Gamma \left(c-a_2\right)}{\Gamma \left(b_1\right) \Gamma \left(c-a_1-a_2\right)} F_{4h} \left(1+a_1+a_2-c,a_1,b_2,a_2-b_2,-a_2+b_2+b_3; 1+a_1-b_1\left| -\frac{1}{z},\frac{1}{x},-\frac{z}{y}\right.\right) \nonumber \\
    &+ (-x)^{-b_1} \frac{ \Gamma \left(a_1-b_1\right) \Gamma \left(c-a_2\right)}{\Gamma \left(a_1\right) \Gamma \left(c-a_2-b_1\right)} F_{4h} \left(1+b_1+a_2-c,b_1,b_2,a_2-b_2,-a_2+b_2+b_3; 1+b_1-a_1\left| -\frac{1}{z},\frac{1}{x},-\frac{z}{y}\right.\right) \nonumber
\end{align}
    
\end{small}
Finally, we consider the third series $S_3$
\begin{align}
    S_3 &= \sum_{m,n,p=0}^\infty \frac{ \left(a_1\right)_m \left(b_1\right)_m \left(b_2\right)_n \left(b_3\right)_p \left(-c+b_2+b_3+1\right)_{-m+n+p}}{ \left(-a_2+b_2+b_3+1\right)_{n+p}} \frac{(-x)^m  y^{-n} z^{-p}}{m! n! p!}\nonumber\\
    &= \sum_{n,p=0}^\infty\frac{\left(b_2\right)_n \left(b_3\right)_p \left(-c+b_2+b_3+1\right)_{n+p} }{\left(-a_2+b_2+b_3+1\right)_{n+p}} \frac{y^{-n} z^{-p} }{n! p! } \, _2F_1\left(a_1,b_1;c-n-p-b_2-b_3|x\right)
\end{align}
Using the analytic continuation of $_2F_1 (\dots ; x)$ around $x=\infty$, we find the analytic continuation of $S_3$
\begin{small}

\begin{align}\label{eqn:LS_infinfinf3}
    S_3' &= (-x)^{-a_1} \frac{ \Gamma \left(b_1-a_1\right) \Gamma \left(c-b_2-b_3\right)}{\Gamma \left(b_1\right) \Gamma \left(c-a_1-b_2-b_3\right)} \sum_{m,n,p=0}^\infty\frac{ \left(a_1\right)_m \left(b_2\right)_n \left(b_3\right)_p \left(-c+a_1+b_2+b_3+1\right)_{m+n+p}}{ \left(a_1-b_1+1\right)_m \left(-a_2+b_2+b_3+1\right)_{n+p}} \frac{x^{-m} y^{-n} z^{-p}}{m! n! p!}\nonumber\\
    &+ (-x)^{-b_1} \frac{ \Gamma \left(a_1-b_1\right) \Gamma \left(c-b_2-b_3\right)}{\Gamma \left(a_1\right) \Gamma \left(c-b_1-b_2-b_3\right)} \sum_{m,n,p=0}^\infty
    \frac{ \left(b_1\right)_m \left(b_2\right)_n \left(b_3\right)_p \left(-c+b_1+b_2+b_3+1\right)_{m+n+p}}{ \left(-a_1+b_1+1\right)_m \left(-a_2+b_2+b_3+1\right)_{n+p}} \frac{x^{-m} y^{-n} z^{-p}}{m! n! p!}\\
     &= (-x)^{-a_1} \frac{ \Gamma \left(b_1-a_1\right) \Gamma \left(c-b_2-b_3\right)}{\Gamma \left(b_1\right) \Gamma \left(c-a_1-b_2-b_3\right)} F_G \left( 1-c+a_1 + b_2+b_3, b_3, b_2, a_1; 1-a_2+b_2+b_3, 1+a_1-b_1\left| \frac{1}{z}, \frac{1}{y}, \frac{1}{x}\right.\right) \nonumber \\
     &+ (-x)^{-b_1} \frac{ \Gamma \left(a_1-b_1\right) \Gamma \left(c-b_2-b_3\right)}{\Gamma \left(a_1\right) \Gamma \left(c-b_1-b_2-b_3\right)}  F_G \left( 1-c+b_1 + b_2+b_3, b_3, b_2, b_1; 1-a_2+b_2+b_3, 1+b_1-a_1\left| \frac{1}{z}, \frac{1}{y}, \frac{1}{x}\right.\right) \nonumber
\end{align}

\end{small}

Thus, the analytic continuation of the $F_S^{(3)}$ around $(\infty,\infty,\infty)$ is sum of the expressions \eeqref{eqn:LS_infinfinf1}, \eeqref{eqn:LS_infinfinf2} and \eeqref{eqn:LS_infinfinf3} multiplied by appropriate prefactors.
\begin{align}
    F_S^{(3)}  &= (-y)^{-a_2} \frac{ \Gamma (c) \Gamma \left(b_2-a_2\right)}{\Gamma \left(b_2\right) \Gamma \left(c-a_2\right)} \times S_1' \nonumber\\
    &+  (-y)^{-b_2}  (-z)^{b_2-a_2} \frac{ \Gamma (c) \Gamma \left(a_2-b_2\right) \Gamma \left(-a_2+b_2+b_3\right)}{\Gamma \left(a_2\right) \Gamma \left(b_3\right) \Gamma \left(c-a_2\right)} \times S_2' \nonumber\\
    &+ (-y)^{-b_2} (-z)^{-b_3}  \frac{ \Gamma (c) \Gamma \left(a_2-b_2-b_3\right)}{\Gamma \left(a_2\right) \Gamma \left(c-b_2-b_3\right)} \times S_3'
\end{align}

We do not write the expression explicitly due to its long length. 
The corresponding domain of convergence can be found using the Horn's theorem as
\begin{align}
    \frac{1}{| x| }<1\land \frac{1}{| x| }+\frac{1}{| y| }<1\land \frac{1}{| y| }<1\land \frac{1}{| x| }+\frac{1}{| z| }<1\land \frac{1}{| z| }<1\land \left| \frac{z}{y}\right| <1
\end{align}
Another analytic continuation of the $F_S^{(3)}$ around $(\infty,\infty,\infty)$ can be readily found by employing the symmetry relation \eeqref{eqn:LS_symmetry}.

It is to be noted that, the analytic continuations above are valid for \textit{generic values} of Pochhammer parameters, which means that the parameters do not assume such values that cause the series to terminate or give divergent results. This typically corresponds to the cases when the difference of parameters is not an integer. However, for the non-generic case, when the difference is an integer, these analytic continuations remain valid as long as a proper limiting procedure is applied. Strategies for using the analytic continuations for non-generic cases have also been implemented in all the four packages developed, which is be discussed in Section \ref{sec:non_generic}.

Due to the large number of analytic continuations derived, it is not possible to present all of them in text form. For the ease of the reader, we have built commands, to be discussed in Section \ref{sec:demo}, so that one can see the explicit form of the analytic continuations directly from the package.
The total number of analytic continuations, including the original series definition of each of the functions, that are stored inside the packages is given below.
\begin{itemize}
    \item \fone: 28 
    \item \fthree: 24 
    \item \lfd: 96
    \item \lfs: 102
\end{itemize}

We emphasize that, the analytic continuations of these functions are valid  for all values of their arguments except for the singular loci of respective functions.


\section{Algorithm of the packages}\label{sec:algorithm}

We now discuss the algorithm used for the package in detail. Three primary issues must be addressed to ensure the efficient evaluation of numerical values within the package, which are as follows.

\begin{enumerate}
    \item \textit{Selection of a suitable analytic continuation}: Although for a point of interest, there are many analytic continuations that can be used for evaluation, not all of them are equally efficient for evaluation purposes because convergence rates vary among them. The analytic continuations, being composed of multiple series, further complicate the matter, as not all series have favourable convergence rates. Hence, a suitable analytic continuation has to be selected for numerical evaluation, taking into account all of these factors.
    
    \item \textit{Strategy to perform the summation efficiently}: With the increase in the fold of the series, the amount of time required to evaluate finite series summation takes an exponentially larger time. Given a $p-$fold series, the number of operations required to perform summation, taking $\texttt{N}$ term in all the summation indices, is $\mathcal{O}(\texttt{N}^{p})$. Hence, any strategy that can reduce the number of operations is the key to achieving faster evaluations. 
    \item \textit{Strategy for numerical evaluation with non-generic parameter values}: In general, the analytic continuations obtained using the method of Olsson are only valid for generic values of the Pochhammer parameters. They, however, are also required for non-generic parameters, which are useful for physical applications as well. For such non-generic parameters, proper limiting procedures have to be used such that the divergences amongst all the series appearing in a certain analytic continuation cancels.
\end{enumerate}
    
We discuss the first two issues below, keeping the last one for Section \ref{sec:non_generic}.

For the first problem, we follow the strategy utilized in \cite{Ananthanarayan:2021bqz}, for the package \ftwo. It relies on defining a certain `rate of convergence' for a given analytic continuation and then selecting the one with the fastest rate of convergence. We have used the same strategy for all the four packages. 

For the second issue, we use two strategies, which we now discuss in detail for both two and three-variable cases. As a demonstrative example, we consider Appell $F_1$, though the same strategies apply to others as well.  
Consider the series representation of $F_{1}$ given in Eq. \eqref{eqn:appellfun}. If we sum over index $n$, we get the following
\begin{equation}\label{eq:f1onesum}
    F_1\left(a, b_1, b_2 ; c | x, y\right) =\sum_{m=0}^{\infty} \frac{(a)_{m}\left(b_1\right)_m}{(c)_{m} m !} x^m \,_{2}F_{1}(a+m,b_{2};c+m|y)
\end{equation}

The strategy now is to use Eq. \eqref{eq:f1onesum}, since the evaluation of $F_{1}$, which originally is given by a double summation, has now been reduced to the evaluation of a single fold summation. For such evaluations, we evaluate the $_{2}F_{1}$ on the right-hand side of Eq. \eqref{eq:f1onesum} using the internal command of \mt, \texttt{Hypergeometric2F1}. 
The time taken for the evaluation using this strategy for various points for Appell $F_{1}$ is compared with that of the usual double summation in Table \ref{table:sum2var}.

\begin{table}[htbp]
\footnotesize
\centering
\begin{tabular}{|c|c|c|c|c|}
\hline
$(x,y)$        & \begin{tabular}[c]{@{}c@{}}Eq. \eqref{eqn:appellfun}\\ ($\texttt{N} = 100$)\end{tabular}                    & \begin{tabular}[c]{@{}c@{}}Eq. \eqref{eq:f1onesum}\\ ($\texttt{N} = 100$)\end{tabular}                   & \begin{tabular}[c]{@{}c@{}}Eq. \eqref{eqn:appellfun}\\ ($\texttt{N} =300$)\end{tabular}                    & \begin{tabular}[c]{@{}c@{}}Eq. \eqref{eq:f1onesum}\\ ($\texttt{N} = 300$)\end{tabular}                   \\ \hline
$(\frac{1}{10},\frac{1}{10})$ & \begin{tabular}[c]{@{}c@{}}1.1927075668481758434\\ t= 0.978377\end{tabular} & \begin{tabular}[c]{@{}c@{}}1.1927075668481758434\\ t= 0.373886\end{tabular} & \begin{tabular}[c]{@{}c@{}}1.1927075668481758434\\ t= 27.0116\end{tabular} & \begin{tabular}[c]{@{}c@{}}1.1927075668481758434\\ t= 0.579974\end{tabular} \\ \hline

$(\frac{1}{10},\frac{9}{10})$ & \begin{tabular}[c]{@{}c@{}}9.6274055869884276609\\ t= 0.731865\end{tabular} & \begin{tabular}[c]{@{}c@{}}9.6274055869884276609\\ t= 0.389833\end{tabular} & \begin{tabular}[c]{@{}c@{}}9.6274915622753831775\\ t= 26.638\end{tabular} & \begin{tabular}[c]{@{}c@{}}9.6274915622753831775\\ t= 0.398074\end{tabular}  \\ \hline

$(\frac{9}{10},\frac{1}{10})$ & \begin{tabular}[c]{@{}c@{}}3.0767526672458511801\\ t= 1.03145\end{tabular} & \begin{tabular}[c]{@{}c@{}}3.0767528856677694923\\ t= 0.469786\end{tabular}  & \begin{tabular}[c]{@{}c@{}}3.0767528856677694748\\ t= 26.7015\end{tabular} & \begin{tabular}[c]{@{}c@{}}3.0767528856677694923\\ t= 1.00073\end{tabular}  \\ \hline

$(\frac{9}{10},\frac{9}{10})$ & \begin{tabular}[c]{@{}c@{}}135.11971472599376751\\ t= 1.05071\end{tabular} & \begin{tabular}[c]{@{}c@{}}135.12030231208480330\\ t= 0.464883\end{tabular} & \begin{tabular}[c]{@{}c@{}}135.12909322010073107\\ t= 27.4262\end{tabular} & \begin{tabular}[c]{@{}c@{}}135.12909322010080844\\ t= 1.18971\end{tabular}  \\ \hline
\end{tabular}
\caption{The values of Appell $F_1$ for various points are obtained using  \eeqref{eqn:appellfun} and  \eeqref{eq:f1onesum}. The Pochhammer parameters are $a=\frac{123}{100},b_1=\frac{234}{100}, b_{2}=\frac{398}{100},c=\frac{47}{10}$. The values are displayed with 20 significant digits. $\texttt{N}$ is the maximum value of each index for finite summation, and $t$ is the time in seconds for the evaluation in a typical run. The run timing corresponds to the time taken for a typical evaluation in a typical laptop.} \label{table:sum2var}
\end{table}

From Table \ref{table:sum2var}, it is observed that Eq. \eqref{eq:f1onesum} outperforms Eq. \eqref{eqn:appellfun} in both time and convergence. The four points are selected to test the performance for points both inside as well as near the boundaries of the convergence region, where the convergence is usually slow. For the point $(9/10,1/10)$, it is observed that Eq. \eqref{eqn:appellfun} is less accurate than Eq. \eqref{eq:f1onesum}, which has already converged for $\texttt{N}= 100$. The point $(9/10,9/10)$ is very close to the boundary (in both directions), and the result is extremely slow to converge even with  \eeqref{eq:f1onesum}. The time taken by  \eeqref{eqn:appellfun} is always higher than  \eeqref{eq:f1onesum}. It is also noticed that the time taken by  \eeqref{eqn:appellfun} increases exponentially when the finite summation limit goes from $\texttt{N}= 100$ to $\texttt{N}= 300$, whereas the increment is not significant with  \eeqref{eq:f1onesum}.

We employ a similar strategy for three variable cases as well. Consider \eeqref{eq:lfd3}, summing over index $p$ we get the following 
\begin{equation}\label{eq:fd3red}
   F_{D}^{(3)}(a,b_{1},b_{2},b_{3};c| x,y,z)= \sum_{m,n=0}^{\infty}\frac{x^m y^n (b_{1})_m (b_{2})_n (a)_{m+n} }{m! n! (c)_{m+n}} \, _2F_1(b_{3},a+m+n;c+m+n|z)
\end{equation}

We now consider the following property of double summation \cite{arfken2011mathematical}
\begin{align}\label{eqn:dsumred}
\sum_{m =0}^{\infty} \sum_{n=0}^{\infty} a (m,n)= \sum_{m=0}^{\infty} \sum_{n=0} ^{m} a(m-n,n) 
\end{align}

In the finite summations where the left-hand side of  \eeqref{eqn:dsumred} has $\texttt{N}$ as the summation limit, $\frac{\texttt{N}^{2}}{2}$ terms are required to be evaluated on the right-hand side, as opposed to $\texttt{N}^{2}$ terms on the left-hand side. Thus, we decrease the number of operators in taking the finite summation by a factor of 2 by rewriting  \eeqref{eq:fd3red} as
\begin{equation}\label{eq:fd3redfinal}
  F_{D}^{(3)}(a,b_{1},b_{2},b_{3}| c;x,y,z)=  \sum_{m=0}^{\infty} \sum_{n=0} ^{m} \frac{y^n (a)_m (b_2)_n x^{m-n} (b_1)_{m-n} \, _2F_1(b_{3},a+m;c+m|z)}{n! (m-n)! (c)_m}
\end{equation}

The time taken for the evaluation with this strategy for various points for  $F_{D}^{(3)}$ is contrasted with that of the usual double summation in Table \ref{table:sum3var}. We observe similar behaviour as table \ref{table:sum2var} for table \ref{table:sum3var} as well.

\begin{table}[htbp]
\footnotesize
\centering
\begin{tabular}{|c|c|c|c|c|}
\hline
$(x,y,z)$           & \begin{tabular}[c]{@{}c@{}}Eq. \eqref{eq:lfd3}\\ (\texttt{N} = 50)\end{tabular}                  & \begin{tabular}[c]{@{}c@{}}Eq. \eqref{eq:fd3redfinal}\\ (\texttt{N} =50)\end{tabular}                   & \begin{tabular}[c]{@{}c@{}}Eq. \eqref{eq:lfd3}\\ (\texttt{N} = 100)\end{tabular}                  & \begin{tabular}[c]{@{}c@{}}Eq. \eqref{eq:fd3redfinal}\\ (\texttt{N} = 100)\end{tabular}                   \\ \hline
$(\frac{1}{10},\frac{1}{10},\frac{1}{10})$ & \begin{tabular}[c]{@{}c@{}}1.0732135550913555676\\ t= 7.26767\end{tabular} & \begin{tabular}[c]{@{}c@{}}1.0732135550913555676\\ t= 0.869634\end{tabular} & \begin{tabular}[c]{@{}c@{}}1.0732135550913555676\\ t= 116.874\end{tabular} & \begin{tabular}[c]{@{}c@{}}1.0732135550913555676\\ t= 1.94078\end{tabular} \\ \hline

$(\frac{1}{10}, \frac{1}{10},\frac{9}{10})$ & \begin{tabular}[c]{@{}c@{}}1.6231078297620145196\\ t= 8.42558\end{tabular} & \begin{tabular}[c]{@{}c@{}}1.6240002995450544771\\ t= 2.00745\end{tabular} & \begin{tabular}[c]{@{}c@{}}1.6239968320342584400\\ t= 124.944\end{tabular} & \begin{tabular}[c]{@{}c@{}}1.6240002995450544771\\ t= 9.87634\end{tabular} \\ \hline

$(\frac{1}{10},\frac{9}{10},\frac{1}{10})$ & \begin{tabular}[c]{@{}c@{}}2.0091501375087473590\\ t= 7.64192\end{tabular} & \begin{tabular}[c]{@{}c@{}}2.0109093110008280114\\ t= 1.7185\end{tabular} & \begin{tabular}[c]{@{}c@{}}2.0109022514155545663\\ t= 125.285\end{tabular} & \begin{tabular}[c]{@{}c@{}}2.0109022330996805423\\ t= 1.87644\end{tabular} \\ \hline

$(\frac{9}{10},\frac{1}{10}, \frac{1}{10})$ & \begin{tabular}[c]{@{}c@{}}2.5054373292884746698\\ t= 8.04876\end{tabular} & \begin{tabular}[c]{@{}c@{}}2.5054313386996026729\\ t= 0.870391\end{tabular} & \begin{tabular}[c]{@{}c@{}}2.5085713622543825408\\ t= 122.186\end{tabular} & \begin{tabular}[c]{@{}c@{}}2.5085713381376995373\\ t= 1.80294\end{tabular} \\ \hline

$(\frac{9}{10},\frac{9}{10}, \frac{9}{10})$ & \begin{tabular}[c]{@{}c@{}}5.6025209788374342344\\ t= 8.30426\end{tabular} & \begin{tabular}[c]{@{}c@{}}5.6011675291597760012\\ t= 2.00511\end{tabular} & \begin{tabular}[c]{@{}c@{}}5.6131499649884968719\\ t= 122.937\end{tabular} & \begin{tabular}[c]{@{}c@{}}5.6131385069711024026\\ t= 9.48653\end{tabular} \\ \hline
\end{tabular}
\caption{Table of values obtained for various points obtained using Eq. \eqref{eq:lfd3} and Eq. \eqref{eq:fd3redfinal}. Pochhammer parameters used are $a=\frac{13}{10},b_{1}=\frac{1}{5},b_2=\frac{1}{7},b_3=\frac{1}{11},c=\frac{11}{13}$. Values are shown up to 20 significant digits, $\texttt{N}$ denotes the maximum value of each of the indices used for finite summation, and $t$ denotes the time taken to complete the evaluation in seconds for a typical run. The run timing corresponds to the time taken for a typical evaluation in a typical laptop.}\label{table:sum3var} 
\end{table}  

It is to be noted that out of many choices of choosing a summation index, to sum over, we choose the one such that the simplest of the hypergeometric functions is obtained. The preference is : $_2F_{1} > \,_{3}F_{2}> \cdots > \,_{p}F_{p-1}$. In the studies of both two variables and three-variable hypergeometric functions, $_{3}F_{2}$ and $_{4}F_{3}$ are rarely obtained, and $_5F_{4}$ or higher order hypergeometric functions are never obtained. This suggests that a particular summed index may be better than the other for numerical evaluation. However, this aspect is not focused on in the present investigation, and it will be a part of future investigation.

\section{Demonstration of the packages} \label{sec:demo}

We now demonstrate the commands of the packages with examples. There are four commands in the packages \lfd and \lfs and the \fone and \fthree contain five commands each, which are enlisted below. 
\begin{itemize}
    \item $\texttt{F1}, \texttt{F3}, \texttt{FD3}, \texttt{FS3}$ : Given a point $(x,y)$ for a two variable functions or $(x,y,z)$ for a three variable functions and the values of the parameters, these commands yield the numerical values of the corresponding functions.
    
    \item $\texttt{F1findall}, \texttt{F3findall}, \texttt{FD3findall}, \texttt{FS3findall}$: Given a point $(x,y)$ or $(x,y,z)$, these commands give a list of all the analytic continuations, valid at that given point. 
      \item $\texttt{F1expose}, \texttt{F3expose}, \texttt{FD3expose}, \texttt{FS3expose}$ : These commands can be used to expose particular analytic continuations that the user wants to see along with its region of convergence.
      
    \item $\texttt{F1evaluate}, \texttt{F3evaluate}, \texttt{FD3evaluate}, \texttt{FS3evaluate}$: Given a point $(x,y)$ for two variable functions or $(x,y,z)$ for three variable functions and the numerical values of the parameters, these commands can be used to evaluate the function of interest using a chosen analytic continuation.
     
      \item $\texttt{F1ROC}, \texttt{F3ROC}$ : These commands are for visual aid to determine where a given point $(x,y)$ lies w.r.t a given analytic continuation. 
\end{itemize}

The packages can be downloaded from the link: 
\begin{center}
\url{https://github.com/souvik5151/Appell_Lauricella_Saran_functions}
\end{center}

Below, we consider the example of the Appell $F_1$ for the sake of demonstration of these commands. The usage of the commands associated with other functions is analogous.
To load the package \fone, we use the following command,

\begin{mmaCell}{Input}
<<AppellF1.wl
\end{mmaCell}
\begin{mmaCell}{Print}
AppellF1 v1.0
Authors : Souvik Bera \& Tanay Pathak

\end{mmaCell}

The main command of the package is the following:

\begin{tcolorbox}

\begin{verbatim}
F1[a, b1, b2, c, x, y, precision, sum_limit, verbose-> True]
\end{verbatim}
    
\end{tcolorbox}

The arguments of the commands are : 
\begin{itemize}
    \item \texttt{a, b1, b2, c} are the Pochhammer parameters and  \texttt{x, y} is the given point $(x,y)$. 
    \item \texttt{precision} refers to the precision up to which the result is required to be displayed.
    \item \texttt{sum\_limit} refers to the upper limit of the finite summation. In all the future text, we will denote it by $\texttt{N}$.
    \item \texttt{verbose} is an option used to display information related to the computation, such as the serial number of the valid analytic continuations inside the package and their corresponding convergence rates, as we show in the example below. By default, the value of this option is chosen to be \texttt{False}.
\end{itemize}

The command can be used as follows. 

\begin{mmaCell}{Input}
F1[1.23, 2.34, 3.98, 4.7, 1.9, .9, 5, 100]
\end{mmaCell}
\begin{mmaCell}{Output}
5.6680 + 17.050 I
\end{mmaCell}

With the value of \texttt{verbose} as \texttt{True}, we get some more information as follows.

\begin{mmaCell}{Input}
F1[1.23, 2.34, 3.98, 4.7, 1.9, .9, 5, 100, verbose-> True]
\end{mmaCell}
\begin{mmaCell}{Print}
valid series : \{9,12,20,23\}
convergence rates :\{\{0.1159109377,23\},\{0.1159355395,9\},\{0.2447218050,20\},
\{0.6068309198,12\}\}
selected series : 23
\end{mmaCell}
\begin{mmaCell}{Output}
5.6680 + 17.050 I
\end{mmaCell}

The command \texttt{F1expose} takes the input in the following way 
\begin{tcolorbox}
\begin{verbatim}
F1expose[series number]
\end{verbatim}
\end{tcolorbox}
The command can be used as follows

\begin{mmaCell}{Input}
F1expose[1]
\end{mmaCell}
\begin{mmaCell}{Output}
\{Abs[x]<1\&\&Abs[y]<1,\mmaFrac{\mmaSup{x}{m} \mmaSup{y}{n} Pochhammer[a,m+n]Pochhammer[b1,m]Pochhammer[b2,n]}{
m!n!Pochhammer[c,m+n]}\}
\end{mmaCell}

The command \texttt{F1findall} takes the input in the following way
\begin{tcolorbox}
\begin{verbatim}
F1findall[{x,y}]
\end{verbatim}
\end{tcolorbox}

As an example,

\begin{mmaCell}{Input}
F1findall[\{1.9, .9\}]
\end{mmaCell}
\begin{mmaCell}{Output}
\{9, 12, 20, 23\}
\end{mmaCell}

The above code implies that the given point \texttt{\{1.9,.9\}} lies inside the region of convergence of analytic continuation number \texttt{\{9, 12, 20, 23\}}. The analytic continuation number corresponds to the number associated with it inside the package internally.

Next, we consider the \texttt{F1evaluate} command. It takes the inputs as
\begin{tcolorbox}
\begin{verbatim}
F1evaluate[#,{a, b1, b2, c, x, y}, precision, sum_limit]    
\end{verbatim}    
\end{tcolorbox}
For instance, $F_1$ is evaluated with the analytic continuation labeled as $\texttt{\#}=\texttt{9}$, with the same set of Pochhammer parameters and $x,y$ as

\begin{mmaCell}{Input}
F1evaluate[9, \{1.23, 2.34, 3.98, 4.7, 1.9, .9\}, 5, 100]
\end{mmaCell}
\begin{mmaCell}{Output}
5.6680 + 17.050 I
\end{mmaCell}

We would like to remark that for better efficiency, the internal evaluation of the finite summation, in both \texttt{F1} and \texttt{F1evaluate} commands (and analogous commands in other packages), is done using \texttt{ParalllelSum} command of \mt. 

All the above commands, which are only demonstrated in the context of the package \fone, have a similar analogue in the other three packages as well. However, in each of the packages \fone and \texttt{AppellF3.wl} there is one more command which is useful for visualization of the region of convergences. The command is \texttt{F1ROC} (and \texttt{F3ROC}). It takes the input as
\begin{tcolorbox}
\begin{verbatim}
F1ROC[{x,y}, #, range]    
\end{verbatim}    
\end{tcolorbox}

For example,

\begin{mmaCell}{Input}
F1ROC[\{1.9, .9\}, 9, \{0, 5\}]
\end{mmaCell}

\begin{figure}[h!]
\centering\includegraphics[width=.3\textwidth]{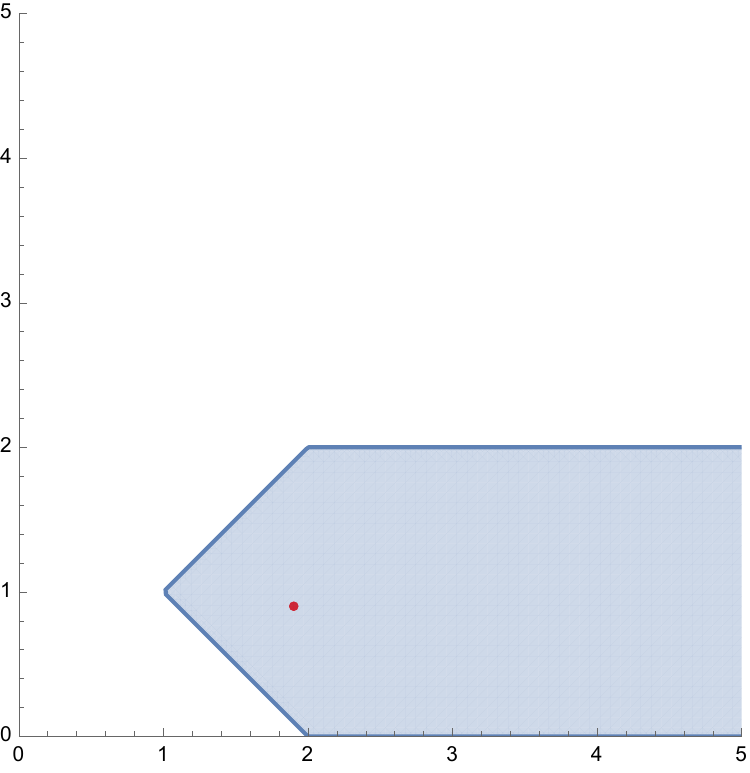}
\caption{Output of \texttt{F1ROC} command.}\label{img:f1roc}
\end{figure}

The output of the above command is the plot shown in Fig.\ref{img:f1roc}. The red dot is the point $(x,y) = (1.9,.9)$, and the blue region is the region of convergence of the analytic continuation labeled as $\texttt{\#} = \texttt{9}$. The range of the plot is $\texttt{range} = [0,5]$, which is specified by the user in the last argument of \texttt{F1ROC} command.

\section{Numerical tests}\label{sec:numtest}
We perform a variety of numerical tests to check the consistency and accuracy of the packages. A detailed list of these numerical checks is given below.

\begin{itemize}
    \item  Consistency check for generic and non-generic cases. These checks ensure that all the valid analytic continuations for a given point gives the same result.
    \item  Test of comparison with the internal commands of \mt and 	\textsc{Maple}.
    \item Testing against reduction formula.
    \item Testing against integral representation.
    \item For the cases of $F_{1}, F_{3}$ and $F_{D}^{(3)}$, checks are also performed by doing the numerical evaluation of Feynman integrals, in which these functions appear. Feynman integrals are evaluated numerically using FIESTA \cite{Smirnov:2021rhf}.
\end{itemize}

\begin{table}[h!]
\centering
\begin{tabular}{|c|c|c|c|}
\hline
$a, b_{1},b_{2},c,x,y$                                                                                                           & \fone                        & \mt                        & 	\textsc{Maple} \\ \hline
\begin{tabular}[c]{@{}c@{}}-4.910115524\\ 0.05551341196\\ 1.272258581\\ 1.701265421\\ 1.865847217\\ -4.593616044\end{tabular}    &   \begin{tabular}[c]{@{}l@{}}$2346.739507$\\ + $0.000021 i$\end{tabular}     &    \begin{tabular}[c]{@{}l@{}}$2346.739507$\\ $+ 0.000021 i$\end{tabular}      &  \# \\ \hline
\begin{tabular}[c]{@{}c@{}}0.06933467465\\ 4.486129287\\ -2.299060382\\ 3.132430057\\ 4.301140034\\ 1.525891559\end{tabular}     &  \begin{tabular}[c]{@{}l@{}}$0.8873796949$\\ $-0.1955529752 i$\end{tabular} &  \begin{tabular}[c]{@{}l@{}}$0.0013530823877$\\ $ - 0.0002148893964 i$ \end{tabular} & \# \\ \hline
\begin{tabular}[c]{@{}c@{}}1.903029939\\ 1.018031546\\ -2.525202153\\ 3.573481224\\ -3.141064731\\ 4.180496216\end{tabular}      & \begin{tabular}[c]{@{}l@{}} $-0.0677462366$ \\ $+ 1.0368642253 i$ \end{tabular}&  \begin{tabular}[c]{@{}l@{}}$-0.0966032877$ \\ $+ 1.0345746851 i$ \end{tabular}    & \begin{tabular}[c]{@{}l@{}}$-0.06774623789$ \\ $+1.036864225\,\mathrm{I}$ \end{tabular}  \\ \hline
\begin{tabular}[c]{@{}c@{}}-0.7885795574\\ 2.712909050\\ -2.552645509\\ -0.1250091986\\ 0.3202377540\\ -1.557083143\end{tabular} & 58.57916734                    & 58.57916734                         & 58.57916733  \\ \hline
\begin{tabular}[c]{@{}c@{}}4.425727083\\ -4.858260819\\ 4.467162437\\ 0.1571461627\\ -0.6971595548\\ -4.125289320\end{tabular}   & -0.002461165757                & -0.01014766698                      & -0.002461165753  \\ \hline
\end{tabular}
\caption{Comparison of numerical values of Appell $F_1$ obtained using the package \fone, inbuilt command \texttt{AppellF1} of \mt and \textsc{Maple}. The symbol \# in the last column means that, \textsc{Maple} fails to give any output in 10 minutes of run time.}\label{table:f1gen}
\end{table}

All the crosschecks are performed in \mt v.13.3 and MAPLE 2021.

In Table \ref{table:f1gen}, we present the comparison of values of Appell $F_1$ obtained using the package \fone, inbuilt command \texttt{AppellF1} of \mt \footnote{These implementations are available in v13.3 and later versions.} and \textsc{Maple}. The Pochhammer parameters and the points $(x,y)$ are chosen randomly in the range $[-5,5]$. In Table \ref{table:f1gen}, we only display the values obtained for five such points, and a \mt notebook containing a larger table of consistency checks is provided in the ancillary file. We observe that there are instances where \textsc{Maple} does not give any output. We also observe that there are instances where results obtained using \mt do not match with the results obtained using the package \fone and \textsc{Maple}, whereas the latter two results are consistent. Furthermore, for all of the mentioned points, we also perform internal consistency checks. Given a random point, we evaluated all the valid analytic continuations and found that values obtained using different analytic continuations are consistent with each other.

\begin{table}[h!]
\centering
\begin{tabular}{|c|c|c|c|}
\hline
$a, b_{1},b_{2},c,x,y$                                                                                                                         & Package                         & \mt                                            & \textsc{Maple} \\ \hline
\begin{tabular}[c]{@{}c@{}}-4.627149094\\ -1.817429336\\ 4.926597166\\ 3.148456748\\ -2.594781226\\ -3.573497861\\ -2.237584115\end{tabular}   & $-4.00373992374711 \times 10^7$ &    \begin{tabular}[c]{@{}l@{}} $3.763432588 \times 10^7$\\ $-1.366207152 \times10^{7} i$\end{tabular}&  \begin{tabular}[c]{@{}l@{}} $ - 4.003739919\times10^{7}$\\ $+  0.002794901917 i$\end{tabular}  \\ \hline
\begin{tabular}[c]{@{}c@{}}1.838637549\\ 0.4850796858\\ 0.09513160021\\ -0.7805808901\\ 1.004405568\\ 2.856794703\\ 3.475873889\end{tabular}   & \begin{tabular}[c]{@{}l@{}}$-0.2910249012$ \\ 
$+ 0.6554323865 i$ \end{tabular}&             \begin{tabular}[c]{@{}l@{}} $-0.2910249012$\\ $+ 0.6554323865 i$\end{tabular}         & \begin{tabular}[c]{@{}l@{}}$- 0.2910249014$ \\ $+ 0.6554323869 i$ \end{tabular}  \\ \hline
\begin{tabular}[c]{@{}c@{}}4.162091546\\ -2.843597171\\ 1.482790024\\ -0.6871600251\\ 0.4147393972\\ 0.8642634792\\ -4.248443060\end{tabular}  & 226772.4287                     & \#                                                      &  226772.4289  \\ \hline
\begin{tabular}[c]{@{}c@{}}3.498347332\\ -2.874934844\\ -2.837329275\\ -1.228710753\\ 2.683926791\\ 0.3173558206\\ -0.3309225764\end{tabular}  & 0.04568785423                   & 0.04568785423                                          &  0.04568785425  \\ \hline
\begin{tabular}[c]{@{}c@{}}1.041216032\\ 2.834971008\\ -0.01534212602\\ 4.704941678\\ 0.5710450502\\ -0.2428307966\\ -2.054034590\end{tabular} & 0.01661851674                   & 0.01661851674                                          &  0.01661851674  \\ \hline
\end{tabular}
\caption{Comparison of numerical values obtained using the package \fthree, inbuilt command of \texttt{AppellF3} of \mt and \textsc{Maple}. The symbol \# in the 3rd column means that \mt fails to give any output in 10 minutes. 
}\label{table:f3gen}
\end{table}

In Table \ref{table:f3gen}, we present a similar comparison of the values obtained using the package \fthree, and inbuilt command \texttt{AppellF3} of \mt and \textsc{Maple}.

\begin{table}[]
\begin{tabular}{|c|cc|cc|}
\hline
\multirow{2}{*}{$a,b_{1},b_{2},c,x,y$}                                                                                                                                 & \multicolumn{2}{c|}{\texttt{AppellF1.wl}}                                                                                                       & \multicolumn{2}{c|}{\textsc{Mathematica}}                                                                                                        \\ \cline{2-5} 
& \multicolumn{1}{c|}{Value}  & Time (s)   & \multicolumn{1}{c|}{Value}                                                                  & Time (s)  \\ \hline
\begin{tabular}[c]{@{}l@{}}-2.683\\ -4.352\\ 0.332\\ -0.729\\ -0.794\\ 0.43\end{tabular}     &  
\multicolumn{1}{c|}{
-55.222948572786781306
} 
& 1.17592  & \multicolumn{1}{c|}{
-55.222948572786781306
}                 & 0.000753 \\ \hline
\begin{tabular}[c]{@{}l@{}}0.708\\ -4.983\\ -1.816\\ 3.181\\ -0.611\\ -0.888\end{tabular} & \multicolumn{1}{c|}{
3.5804485832798687762
}                  & 1.39072  & \multicolumn{1}{c|}{
3.5804485832798687762
}                  & 0.000741 \\ \hline
\begin{tabular}[c]{@{}l@{}}-4.291\\ -1.58\\ -1.666\\ 2.573\\ 0.951\\-0.868\end{tabular}   & \multicolumn{1}{c|}{
-0.89294510000656561514
}                & 0.963402 & \multicolumn{1}{c|}{
-2.0932973267787984230
}                  & 0.000725 \\ \hline
\begin{tabular}[c]{@{}l@{}}-0.343\\ 2.549\\ 2.916\\ 1.11\\ 0.602\\ -0.036\end{tabular}       & \multicolumn{1}{c|}{
0.26120389064476188148
}                 & 0.946174 & \multicolumn{1}{c|}{
0.26120389064476188148
}                 & 0.00072  \\ \hline
\begin{tabular}[c]{@{}l@{}}-0.233\\ 4.274\\ 0.116\\ -4.856\\ 0.956\\0.608\end{tabular}        & \multicolumn{1}{c|}{
$5.7807919398552172706 \times 10^{13}$
} & 1.7405   & \multicolumn{1}{c|}{
$5.7807919398552172706 \times 10^{13}$
} & 0.000788 \\ \hline
\end{tabular}
\caption{Table of comparison of \texttt{AppellF1.wl} package with \textsc{Mathematica} in the original domain of convergence. To obtain the results from the package we take $\texttt{N}= 200$.}\label{table:f1packageoriginal}
\end{table}

\begin{table}[]
\begin{tabular}{|c|cc|cc|}
\hline
\multirow{2}{*}{$a,b_{1},b_{2},
c_1, c_2,x,y$}                                                                  & \multicolumn{2}{c|}{\texttt{AppellF3.wl}}                                                                                       & \multicolumn{2}{c|}{\textsc{Mathematica}}                                                      \\ \cline{2-5}  & \multicolumn{1}{c|}{Value} & Time (s) & \multicolumn{1}{c|}{Value} & Time (s) \\ \hline
\begin{tabular}[c]{@{}l@{}}-3.333\\  -3.26\\  -0.897\\ -2.175\\  1.013 \\ -0.383\\  0.47\end{tabular}   & \multicolumn{1}{c|}{
1.8665976342450988885
}  & 1.40378  & \multicolumn{1}{c|}{
1.8665976342450988885
}  & 0.000123 \\ \hline
\begin{tabular}[c]{@{}l@{}}-0.62\\ 1.719\\  3.128\\ -3.486\\  3.452\\  -0.02\\  -0.456\end{tabular}     & \multicolumn{1}{c|}{
2.1297626615431425458
}  & 1.33443  & \multicolumn{1}{c|}{
2.1297626615431425458
}  & 0.000041 \\ \hline
\begin{tabular}[c]{@{}l@{}}-3.695\\  0.826\\  3.312\\  3.008\\  3.637\\  0.351\\  0.931\end{tabular}    & \multicolumn{1}{c|}{
2.4380309068841122890
}  & 1.34582  & \multicolumn{1}{c|}{
2.4380309068841122890
}  & 0.000132 \\ \hline
\begin{tabular}[c]{@{}l@{}}-3.362\\  -0.871\\ -3.907\\  -3.372\\  -3.132\\  0.123\\  0.424\end{tabular} & \multicolumn{1}{c|}{
0.43899084581055544234
} & 1.65237  & \multicolumn{1}{c|}{
0.43899084581055544234
} & 0.000066 \\ \hline
\begin{tabular}[c]{@{}l@{}}0.312\\  0.164\\ -2.193\\  -3.313\\ 4.326\\  -0.262\\0.643\end{tabular}              & \multicolumn{1}{l|}{
0.97171986750193850061
} & 2.75102  & \multicolumn{1}{l|}{$0.971719867501939^{*}$} & -        \\ \hline
\end{tabular}
\caption{Table of comparison of \texttt{AppellF3.wl} package with \textsc{Mathematica} in the original domain of convergence. To obtain the results from the package we take $\texttt{N}= 200$. The symbol * in the last result means that we cannot more precise results for this case using \mt.}\label{table:f3packageoriginal}
\end{table}

Results of various tests concerning the error analysis of the results obtained using the packages are provided in Appendix \ref{appendix:error}. 
As a simple additional check, we compare the values obtained using \mt and the two packages \fone and \fthree in the original domain of convergence of the corresponding series $F_{1}$ and $F_{3}$ respectively. The result for the same are shown in Table \ref{table:f1packageoriginal} and \ref{table:f3packageoriginal}. We observe that for this case where the results can be obtained by simply summing the corresponding series representation of these functions, \mt at times gives incorrect results and sometimes fails to give results within a certain time scale relevant for practical purposes. The incorrect result entry is the entry 3 of Table \ref{table:f1packageoriginal}. In entry 5 of Table \ref{table:f3packageoriginal} we observe that \mt is not able to give a result with more precision than shown in the table. As an additional test for this particular case, we substitute the numerical values of parameters with exact rational numbers. We notice that with the rational numbers as arguments, \mt does not yield output within practical times. However, in other cases, we observe that the time taken by \mt is significantly lower as compared to packages.

\newpage


\section{Numerical test with non-generic values of parameters : An application to Feynman integrals}\label{sec:application}

The numerical test of our packages for the non-generic values of Pochhammer parameters is discussed in this Section. It is frequently observed that the evaluation of Feynman integral in hypergeometric function representation produces function with non-generic values of Pochhammer parameters. Thus, our tests of these packages for non-generic values of parameters are motivated by Feynman integrals. 


\subsection{Passarino-Veltman functions}

 Consider the one loop $N$-point scalar Feynman integral, for the case when all external momenta are zero, $p_{1} = p_{2} = \cdots = p_{N} =0$. Each propagator has index $\nu_{i}$ and mass $m_{i}$. The corresponding integral is given below.
\begin{equation}
    I^{N} ( \{\nu_{i}\};\{m_{i}\}) = \int \frac{d^{d}k}{(k^{2}-m_{1}^{2})^{\nu_{1}} \cdots (k^{2}-m_{N}^{2})^{\nu_{N}}}
\end{equation}

Note that, for integer values of powers of the propagators, this integral can be evaluated using partial fractioning. For generic values of powers of propagators, 
the integral can be expressed in terms of $(N-1)$ variable Lauricella $F_{D}^{(N-1)}$ function \cite{Davydychev:1990cq}. 
\begin{align}
  I^{N} ( \{\nu_{i}\};\{m_{i}\}) &= \pi^{d/2} i^{-d} (-m_{N})^{d/2- \sum_{j} \nu_{j}} \frac{\Gamma(\sum_{j} \nu_{j}-d/2)}{\Gamma(\sum_{j} \nu_{j})}  \nonumber \\
& \times  F_{D}^{(N-1)} \left( \sum_{j} \nu_{j}-\frac{d}{2},\nu_{1},\cdots,\nu_{N-1};\sum_{j} \nu_{j} \Big| 1-\frac{m_{1}^{2}}{m_{N}^{2}},\cdots,1-\frac{m_{N-1}^{2}}{m_{N}^{2}}  \right)
\end{align}

Let us now consider some particular situations. When $N=3$, we have the Passarino-Veltman $C_0$ function 
\begin{align}\label{eqn:3ptF1}
    C_0 :&=  I^3 (\{1,1,1\}, \{m_1, m_2, m_3\}) \nonumber\\
    &=  \frac{1}{2} i^{2 \varepsilon +1} \pi ^{2-\varepsilon } \Gamma (\varepsilon +1) \left(-m_3^2\right){}^{-\varepsilon -1} F_1\left(\varepsilon +1,1,1;3 \Big| 1-\frac{m_1^2}{m_3^2},1-\frac{m_2^2}{m_3^2}\right)
\end{align}
and for $N=4$,
\begin{align}\label{eqn:4ptFD}
    D_0 :&= I^3 (\{1,1,1,1\}, \{m_1, m_2, m_3,m_4\}) \nonumber\\
    &= \frac{1}{6} i^{2 \varepsilon +1} \pi ^{2-\varepsilon } \Gamma (\varepsilon +2) \left(-m_4^2\right){}^{-\varepsilon -2} F_D^{(3)}\left(\varepsilon +2,1,1,1;4 \Big| 1-\frac{m_1^2}{m_4^2},1-\frac{m_2^2}{m_4^2},1-\frac{m_3^2}{m_4^2}\right)
\end{align}

It is well-known, and in fact obvious, from the expression of Eqs. \eqref{eqn:3ptF1} and \eqref{eqn:4ptFD} that, the 3- and 4-point functions are convergent in $d = 4$ (i.e., $\varepsilon = 0$)
\begin{align}\label{eqn:c0at4dim}
    C_0 (d=4) = -\frac{i \pi ^2 }{2 m_3^2}F_1\left(1,1,1;3 \Big| 1-\frac{m_1^2}{m_3^2},1-\frac{m_2^2}{m_3^2}\right)
\end{align}
valid for $\left| 1-\frac{m_1^2}{m_3^2}\right| <1\land \left| 1-\frac{m_2^2}{m_3^2}\right| <1$,
and 
\begin{align}\label{eqn:D0at4dim}
    D_0 (d=4) = \frac{i \pi ^2 }{6 m_4^4} F_D^{(3)}\left(2,1,1,1;4 \Big| 1-\frac{m_1^2}{m_4^2},1-\frac{m_2^2}{m_4^2},1-\frac{m_3^2}{m_4^2}\right)
\end{align}
valid for $\left| 1-\frac{m_1^2}{m_4^2}\right| <1\land \left| 1-\frac{m_2^2}{m_4^2}\right| <1\land \left| 1-\frac{m_3^2}{m_4^2}\right| <1$.

It is to be noted that, the $\varepsilon$-expansion of Appell $F_1$ and Lauricella $F_D^{(3)}$ functions in Eqs \eqref{eqn:3ptF1} and \eqref{eqn:4ptFD}  can be found using computer programs such as Xsummer \cite{Moch:2005uc}, nestedsums \cite{McLeod:2020dxg}, \texttt{MultiHypExp} \cite{Bera:2023pyz} and Diogenes \cite{Bezuglov:2023owj}. Moreover, there exists  reduction formulae of Appell $F_1$ and Lauricella $F_D^{(3)}$ functions, for integer valued Pochhammer parameters, to simpler functions such as ordinary logarithms. These are available in the literature and can also be obtained using the \texttt{ReduceFunction} command of \texttt{MultiHypExp} package. We find,
\begin{align}\label{eq:f1red}
    F_1(1,1,1;3 \mid x,y) = \frac{2 \log (1-x)}{x (x-y)}+\frac{2 \log (1-y)}{x-y}-\frac{2 \log (1-x)}{x-y}-\frac{2 \log (1-y)}{y (x-y)}
\end{align}
and 
\begin{align}\label{eq:fdred}
    F_D^{(3)}(2,1,1,1;4 \mid x,y,z) 
    &= \frac{6 \log (1-x)}{x (x-y) (x-z)}+\frac{6 \log (1-y)}{(x-y) (y-z)}+\frac{6 \log (1-z)}{(x-z) (z-y)}\nonumber\\
    &-\frac{6 \log (1-x)}{(x-y) (x-z)}-\frac{6 \log (1-y)}{y (x-y) (y-z)}-\frac{6 \log (1-z)}{z (x-z) (z-y)}
\end{align}
Following the pattern, one can easily generalize the reduction formula for $F_D^{(N)}$.

Clearly, one needs to find analytic continuations of the Appell $F_1$ and Lauricella $F_D^{(3)}$  to find the numerical values of the corresponding integrals beyond the respective domain of convergences. We use our packages to find the numerical value of these functions and compare with \texttt{FIESTA5} \cite{Smirnov:2021rhf} and the inbuilt \texttt{AppellF1} command of \mt and \textsc{Maple}. The comparisons are reported in Tables \ref{table:f1red} and \ref{table:fdred}.

\begin{table}[h!]
\scriptsize
\centering
\begin{tabular}{|c|c|c|c|c|c|}
\hline
$\{ m_{1}^{2}, m_{2}^{2}, m_{3}^{2}\}$ & \texttt{FIESTA5} & Eq. \eqref{eq:f1red} & \fone & \mt& \textsc{Maple} \\ \hline
 \{1,4,9\}                                 &     \begin{tabular}[c]{@{}l@{}}0.124697034815251\\ $\pm 7.55843197 \times10^{-7}$\end{tabular}

     &   0.124697033602012     &     0.124697033602012    &      0.124697033602012       &     0.124697033602013 \\ \hline
\{9,4,1\}                                  &     \begin{tabular}[c]{@{}l@{}}0.124697034815251\\ $\pm 7.55843197 1 \times 10^{-7}$\end{tabular}

     &    0.124697033602012    &    0.124697033602012     &       0.124697033602012      &   0.124697033602012    \\ \hline
   \{0,4,9\}                               &   \begin{tabular}[c]{@{}l@{}} 0.16218604696104502\\ $\pm 8.34593085 \times 10^{-7}$\end{tabular}

       &    0.162186043243266    &  0.162186043243266       &       0.162186043243266      &    0.162186043243266   \\ \hline
 \{$\frac{1}{25}, \frac{49}{64}, \frac{1}{169}$\}                                 &    \begin{tabular}[c]{@{}l@{}} $3.662930006248521$ \\ $\pm 0.000023800445347$\end{tabular}      &      3.66292995929945  &   3.66292995929945      &      3.66292995929945       &   3.66292995929945    \\ \hline
\end{tabular}
\caption{In this Table, we evaluate the $C_0$ function in \eeqref{eqn:c0at4dim} with four different sets of values of masses using our package \fone (with $\texttt{N} = 250$) and compare the outputs against the  inbuilt command \texttt{AppellF1} of \mt and \textsc{Maple}, reduction formula of Appell $F_1$ (\eeqref{eq:f1red}) and the results from \texttt{FIESTA5}. Note that, to be consistent with the definition of Feynman integrals of \texttt{FIESTA5}, we multiply a factor of $-\frac{1}{i \pi^2}$ with \eeqref{eqn:c0at4dim}.
}\label{table:f1red}
\end{table}

\begin{table}[h!]
\centering
\begin{tabular}{|c|c|c|c|}
\hline
$m_{1}^{2}, m_{2}^{2}, m_{3}^{2},m_{4}^{2}$ & \texttt{FIESTA5} & Eq. \eqref{eq:fdred} & \lfd  \\ \hline
 \{1,4,9,16\}                                 &      \begin{tabular}[c]{@{}l@{}}0.004611058631625\\ $\pm  3.6492129\times 10^{-8}$\end{tabular}

     &   0.00461105850390750     &       0.00461105850390750   \\ \hline
   \{9,16,1,4\}                                &    \begin{tabular}[c]{@{}l@{}}0.004611058631625\\ $\pm  3.6492129\times 10^{-8}$\end{tabular}      &  0.00461105850390750      &  \begin{tabular}[c]{@{}l@{}}0.00461105850390750\\ $+5.150233 \times 10^{-11} i$\end{tabular}      \\ \hline
   \{0,4,9,16\}                                &      \begin{tabular}[c]{@{}l@{}}0.006665930571901001\\ $\pm  3.0497681\times 10^{-8}$\end{tabular}      &    0.00666593044999165    &       0.00666593044999165   \\ \hline
 $\left\{\frac{1}{100},\frac{9}{49},9,\frac{16}{9}\right\}$                                 &       \begin{tabular}[c]{@{}l@{}}0.126545889960494\\ $\pm   2.108113451\times 10^{-6}$\end{tabular}   &   0.126545880468854     &  0.126545880468854      \\ \hline
\end{tabular}
\caption{
In this Table, we evaluate the $D_0$ function in \eeqref{eqn:D0at4dim} with four different sets of values of masses using our package \lfd (with $\texttt{N} = 300$) and compare the outputs against the reduction formula of Lauricella $F_D^{(3)}$ (\eeqref{eq:fdred}) and the results from \texttt{FIESTA5}. Note that, to be consistent with the definition of Feynman integrals of \texttt{FIESTA5}, we multiply a factor of $\frac{1}{i \pi^2}$ with \eeqref{eqn:D0at4dim}.
}\label{table:fdred}
\end{table}

\subsection{Photon-photon scattering}
We now consider another physical example where the Appell $F_{3}$ function appears, the case of photon-photon scattering \cite{Davydychev:1993ut}. The loop amplitude for the four-point function can be written as follows
\begin{equation}\label{eq:photonf3}
    A(s,t; m) = \frac{i \pi^{2}}{6 m^{4}} F_{3}\left(1,1,1,1;\frac{5}{2} \Big| \frac{s}{4 m^{2}}, \frac{t}{4 m^{2}}\right)
\end{equation}
where $s$ and $t$ are the Mandelstam variables and $m$ is the mass of the propagators. 

The right hand side of Eq. \eqref{eq:photonf3}, is further given by the following reduction formula \cite{Davydychev:1993ut} 
\begin{align} \label{eq:redf3}
F_3(1,1,& 1,1 ; 5 / 2 \mid x, y) \nonumber \\
&= \frac{3}{4 x y \beta_{x y}}\left\{2 \ln ^2\left(\frac{\beta_{x y}+\beta_x}{\beta_{x y}+\beta_y}\right)+\ln \left(\frac{\beta_{x y}-\beta_x}{\beta_{x y}+\beta_x}\right) \ln \left(\frac{\beta_{x y}-\beta_y}{\beta_{x y}+\beta_y}\right)-\frac{\pi^2}{2}\right. \nonumber \\
& \left.\quad+\sum_{i=x, y}\left[2 \operatorname{Li}_2\left(\frac{\beta_i-1}{\beta_{x y}+\beta_i}\right)-2 \operatorname{Li}_2\left(-\frac{\beta_{x y}-\beta_i}{\beta_i+1}\right)-\ln ^2\left(\frac{\beta_i+1}{\beta_{x y}+\beta_i}\right)\right]\right\},
\end{align}
where
$\beta_x \equiv \sqrt{1-\frac{1}{x}}, \quad \beta_y \equiv \sqrt{1-\frac{1}{y}}, \quad \beta_{x y} \equiv \sqrt{1-\frac{1}{x}-\frac{1}{y}} .
$.

\begin{table}[h!]
\centering
\scriptsize
\begin{tabular}{|c|c|c|c|c|c|}
\hline
$\{ $s,t,m$\}$ & \texttt{FIESTA5} & Eq. \eqref{eq:redf3} & \fthree & \mt& \textsc{Maple} \\ \hline
$\{ $1,1,1$\}$ & \begin{tabular}[c]{@{}l@{}}0.207247806735362\\ $\pm 2.069729919\times 10^{-6}$\end{tabular}  & 0.207247795188039 & 0.207247795188039 &  0.207247795188039& 0.207247795188039\\ \hline

$\{ $10,2,1$\}$ & \begin{tabular}[c]{@{}l@{}}0.09094036900538302\\ $ +0.307804156595376 i$\\
$\pm (0.000042877912657$\\
$+0.000037404295047 i)$
\end{tabular}  & 
\begin{tabular}[c]{@{}l@{}}0.090940335591268\\ $-0.307803818883125 i$\end{tabular} 
& \begin{tabular}[c]{@{}l@{}}0.090940335591268\\ $-0.307803818883124 i$\end{tabular} & * & \begin{tabular}[c]{@{}l@{}}0.0909403355912683\\ $- 0.307803818883126 i$\end{tabular}\\ \hline

$\{ $5,10,1$\}$ & \begin{tabular}[c]{@{}l@{}}0.37601608623542404\\ $ +1.029892882655468 i$\\
$\pm (0.000125266562509$\\
$+0.000105570122816 i)$
\end{tabular}  & 

\begin{tabular}[c]{@{}l@{}}0.376014441580313\\ $-1.029891629753789  i$\end{tabular}& \begin{tabular}[c]{@{}l@{}}0.376014441580316\\ $-1.029891629753789 i$\end{tabular} & * & \begin{tabular}[c]{@{}l@{}}0.376014441582954\\ $-1.02989162975635 i $\end{tabular} \\ \hline

$\{ $11,10,1$\}$ & \begin{tabular}[c]{@{}l@{}}-0.5345491671227911\\ $ -0.339676045083691 i$\\
$\pm (0.000039238248217$\\
$+0.000033382979669 i)$
\end{tabular}  & 
\begin{tabular}[c]{@{}l@{}}-0.534549424240674\\ $+0.339676241448472 i$\end{tabular}
& \begin{tabular}[c]{@{}l@{}}-0.534549424240674\\ $+0.339676241448473 i$\end{tabular} & * & \begin{tabular}[c]{@{}l@{}}$-0.534549424241419$\\ $ +0.339676241449579 i $\end{tabular} \\ \hline
\end{tabular}
\caption{ The results from the reduction formula (\eeqref{eq:redf3}) are obtained with a small negative imaginary part in order to avoid the evaluation on the branch cuts. The numbers in the 4th column are produced using the package \fthree with $20$ precision and with  $\texttt{N}$ = 200. The `*' in the 5th column denotes that \mt is unable to give any output.  
}\label{table:f3red}
\end{table}

In Table \ref{table:f3red}, we compare the values obtained using the reduction formula of $F_3$, (i.e., Eq. \eqref{eq:redf3}), the package \fthree and using the inbuilt \texttt{AppellF3} commands of \mt and \textsc{Maple} for four kinematic points. Also, the result of the Feynman integral using \texttt{FIESTA5} are provided in the second column. Note that, there is a mismatch in the sign of the imaginary parts of the numbers from the \texttt{FIESTA5} column with the corresponding numbers from the other columns. This is due to the different conventions used to evaluate the functions on the branch cut. From the Feynman's $- i \varepsilon$, we observe that, the amplitude \eeqref{eq:photonf3} has to be evaluated as

\begin{align}\label{eq:ieps}
    A(s,t; m) &= \frac{i \pi^{2}}{6 m^{4}} F_{3}\left(1,1,1,1;\frac{5}{2} \Big| \frac{s}{4 m^{2} - i \varepsilon}, \frac{t}{4 m^{2} - i \varepsilon}\right)\\
    &= \frac{i \pi^{2}}{6 m^{4}} F_{3}\left(1,1,1,1;\frac{5}{2} \Big| \frac{s}{4 m^{2} } + i \varepsilon, \frac{t}{4 m^{2} } + i \varepsilon\right)
\end{align}
whereas, in the mathematics literature and in the implementation of the package \fthree, on the branch cut, the Appell $F_3$ is evaluated as
\begin{align}
    \lim_{\varepsilon \rightarrow 0} F_3(a_1,a_2,b_1,b_2;c| x- i \varepsilon,y - i \varepsilon)
\end{align}
which is the reason behind the mismatch of the sign of the imaginary part of the numbers from the second column to the others in Table \ref{table:f3red}. We have also confirmed this observation by evaluating the reduction formula \eeqref{eq:redf3} and the derived analytic continuations using the $+i \varepsilon$ convention and found it to be consistent with \texttt{FIESTA5}.

\subsection{Lauricella Saran $F_{S}^{(3)}$}
As a simple test of our package, we  now consider a reduction formula taken from \cite{Bera:2023pyz},

\begin{small}

\begin{equation}\label{eq:redfs}
 F_{S}(1,1,1,1,1;2\mid x,y,z) =   -\frac{x \log (1-x)}{(x (y-1)-y) (x (z-1)-z)}+\frac{y \log (1-y)}{(x (y-1)-y) (y-z)}-\frac{z \log (1-z)}{(x (z-1)-z) (y-z)}
\end{equation}
    
\end{small}

The comparison is reported in Table \ref{table:fsred}.

\begin{table}[h]
\centering
\begin{tabular}{|c|c|c|}
\hline
$(x,y,z)$                                                       & Eq. \eqref{eq:redfs}                                         & \lfs                                                             \\ \hline
\begin{tabular}[c]{@{}c@{}}-3.400643617\\ 0.7907915243\\
2.852959631\end{tabular}   & \begin{tabular}[c]{@{}c@{}}$-0.7166971853$\\ $-0.4747886477 i$\end{tabular}       & \begin{tabular}[c]{@{}c@{}}$-0.7166971853$\\ $-0.4747886477 i $\end{tabular}       \\ \hline

\begin{tabular}[c]{@{}c@{}}2.863643067\\ 3.310633871\\ 4.187269367\end{tabular}  & \begin{tabular}[c]{@{}c@{}}$0.0550038448$ \\ $-1.1016637066 i $\end{tabular}  & \begin{tabular}[c]{@{}c@{}}$0.0550038448$ \\ $-1.1016637066 i $\end{tabular}  \\ \hline

\begin{tabular}[c]{@{}c@{}}3.217878146\\ 1.913011642\\ 3.741614131\end{tabular}  & \begin{tabular}[c]{@{}c@{}}$0.006843860 $\\ $-3.882712970 i$\end{tabular}  & \begin{tabular}[c]{@{}c@{}}$0.006843860 $\\ $-3.882712970 i$\end{tabular}  \\ \hline

\begin{tabular}[c]{@{}c@{}}4.149336056\\ -0.2628593004\\
1.966914729\end{tabular} & \begin{tabular}[c]{@{}c@{}}$0.447598037 $\\ $+2.635667205 i $\end{tabular} & \begin{tabular}[c]{@{}c@{}}$0.447598037 $\\ $+2.635667205 i $\end{tabular} \\ \hline

\begin{tabular}[c]{@{}c@{}}1.410789108\\ 4.730065747\\
0.7786036673\end{tabular}  & \begin{tabular}[c]{@{}c@{}}$ 0.52671335 $\\ $+14.69790065 i $\end{tabular}   & \begin{tabular}[c]{@{}c@{}}$ 0.52671335 $\\ $+14.69790065 i $\end{tabular}   \\ \hline
\end{tabular}
\caption{Table of comparison of values of $F_{S}^{(3)}(1,1,1,1,1;2|x,y,z)$ obtained using \lfs (with $\texttt{N}=200$) are compared with the value obtained using reduction formula Eq.\eqref{eq:redfs}. }\label{table:fsred}
\end{table}

\subsection{A note on evaluation with non-generic parameters} \label{sec:non_generic}

In principle, the analytic continuations derived using the method of Olsson are only valid for generic values of Pochhammer parameters. In other words, we tacitly assume that, the parameters do not take such values which results either in the termination of the series or yielding divergent results. However, these analytic continuations can also be used for numerical evaluation of functions with non-generic values of parameters as we show below.

For the sake of demonstration, let us take the example of Lauricella $F_D^{(3)}(2,1,1,1;4;x,y,z)$ with the value of arguments corresponding to the second row of Table \ref{table:fdred}.
\begin{align*}
    (x,y,z) =\left(1-\frac{m_1^2}{m_4^2},1-\frac{m_2^2}{m_4^2},1-\frac{m_3^2}{m_4^2}\right) = \left(-\frac{5}{4},-3,\frac{3}{4}\right)
\end{align*}
We also consider one particular analytic continuation (say, \# 92) which is valid at our point of interest and given by
\begin{align*}
    \text{FD3\#92} &= (1-x)^{-b_1} (1-y)^{-b_2} (1-z)^{-b_3} \left(-\frac{z-1}{z}\right)^{b_3} \frac{\Gamma (c) \Gamma \left(a-b_2\right) \Gamma \left(-a+c-b_3\right)}{\Gamma (a) \Gamma (c-a) \Gamma \left(c-b_2-b_3\right)}\nonumber\\
    &\times \sum_{m,n,p = 0}^\infty \frac{\left(b_1\right)_m \left(b_2\right)_n \left(b_3\right)_p \left(-a+c-b_3\right)_{m+n-p}}{m! n! p! \left(-a+b_2+1\right)_n \left(c-b_2-b_3\right)_{m-p}} \left(\frac{x}{x-1}\right)^m \left(\frac{1}{1-y}\right)^n \left(\frac{z-1}{z}\right)^p\\
    &+ (1-x)^{-b_1} (1-y)^{-b_2} (1-z)^{-b_3}  \left(-\frac{z-1}{z}\right)^{c-a}\frac{ \Gamma (c) \Gamma \left(a-c+b_3\right)}{\Gamma (a) \Gamma \left(b_3\right)}\\
    &\times \sum_{m,n,p = 0}^\infty \frac{ \left(b_1\right)_m \left(b_2\right)_n \left(-a+b_2+1\right)_{n+p} (c-a)_{m+n+p}}{m! n! p! \left(-a+b_2+1\right)_n \left(-a+c-b_3+1\right)_{m+n+p}} \left(\frac{x (z-1)}{(x-1) z}\right)^m \left(\frac{z-1}{(y-1) z}\right)^n  \left(\frac{z-1}{z}\right)^p\\
    &+(1-y)^{-a} (1-x)^{-b_1} (1-z)^{-b_3} \left(-\frac{z-1}{z}\right)^{b_3}\frac{ \Gamma (c) \Gamma \left(b_2-a\right)}{\Gamma \left(b_2\right) \Gamma (c-a)}\\
    &\times \sum_{m,n,p = 0}^\infty \frac{ (a)_n \left(b_1\right)_m \left(b_3\right)_p \left(c-b_2-b_3\right)_{m+n-p}}{m! n! p! \left(a-b_2+1\right)_n \left(c-b_2-b_3\right)_{m-p}}  \left(\frac{x}{x-1}\right)^m \left(\frac{1}{1-y}\right)^n \left(\frac{z-1}{z}\right)^p\\
    & =  \sum_{i = 1,2,3} \text{prefactor}_i \times \text{series}_i
\end{align*}

We observe that, for the particular values of parameters
\begin{align}
    a = 2, b_1 = 1, b_2 = 1, b_3 = 1, c= 4
\end{align}
the $\text{prefactor}_2$ and $\text{prefactor}_3$ are divergent and $\text{prefactor}_1$ is finite. Indeed, by setting 
\begin{align}
    a = 2, b_1 = 1, b_2 = 1 + \varepsilon, b_3 = 1, c= 4 -\varepsilon
\end{align}
we find that,
\begin{align}
    \text{prefactor}_2 &= \frac{1}{\varepsilon} \frac{6 (z-1)}{ (x-1) (y-1) z^2} + O(\varepsilon^0) \hspace{.3cm} \text{and ~~ series}_2 = O(\varepsilon^0) \\
    \text{prefactor}_3 &= \frac{1}{\varepsilon} \frac{6}{ (x-1) (y-1)^2 z} + O(\varepsilon^0)\hspace{.3cm} \text{and ~~ series}_3 = O(\varepsilon^0)
\end{align}
From the definition, one expects that $F_D^{(3)}(2,1,1,1;4|x,y,z)$ to be finite. Thus,
\begin{align}
    \sum_{i = 2,3} \text{prefactor}_i \times \text{series}_i = \text{finite}
\end{align}

The divergent parts coming from two different terms cancel out eventually. Numerically, to get an accurate cancellation, the series has to be summed over a large number of terms $\texttt{N}$.

Inside the packages, when dealing with non-generic parameter values, the analytic continuations are evaluated with a shift of $q \varepsilon$ (where $q$ is a rational number and $\varepsilon$ being a very small imaginary number), to the given values of parameters. In such cases, to get more accurate results, the users are advised to take the summation with the higher number of terms.
The impact of the variation of the number of terms in the summation on the evaluation of $F_D^{(3)}(2,1,1,1;4|x,y,z)$ is elucidated in Table \ref{table:varyingterms}, showcasing how the result is contingent upon the number of terms employed for summation.

\begin{table}[h!]
\centering
\begin{tabular}{|c|l|c|}
\hline
 $\texttt{N}$ & Results from  \lfd & Expected value from \eeqref{eq:fdred}\\ \hline
    10 &     $0.40027651208478822796+8.4411689391473037971\times 10^{18} i$        &       \\ \hline
    50&   $0.44265212069383465865+1.4798709714436027997\times 10^{15} i$          &       \\ \hline
    100&  $0.44266161616971901410+2.9853737369244941467\times 10^{10} i$           &       \\ \hline    
    150&     $0.44266161637511554096+602245.50120234373334 i$        &       \\ \hline
200&     $0.44266161637511983785+12.149220689939886743 i$        &       \\ \hline
    250&    $0.44266161637511983794+0.00024508869402624438466 i$         &       \\ \hline
        300&    $0.44266161637511983794+4.9442239525071142523\times 10^{-9} i$         &       \\ \hline
            350&   0.44266161637511983794          &       \\ \hline
                400&     0.44266161637511983794        &    0.44266161637511983794   \\ \hline
\end{tabular}
\caption{The variation of the results with varying numbers of terms employed for summation is shown. The command \texttt{FD3evaluate[92,\{2,1,1,1,4,-(5/4),-3,3/4\},20, \texttt{N}]} is used to generate the values of $F_D^{(3)}$ using the analytic continuation numbered \#92 with 20 precision and shown in the second column. The expected value in the last column is obtained using the reduction formula \eeqref{eq:fdred}. }\label{table:varyingterms}
\end{table}

\section{Summary \& Outlook} \label{sec:summary}


A detailed investigation of various two and three-variable hypergeometric series, viz. Appell $F_{1}$ and $F_{3}$, Lauricella $F_{D}^{(3)}$ and Lauricella-Saran $F_{S}^{(3)}$, is carried out in this work, and their analytic continuations are obtained using the method of Olsson. 
Four packages, named \fone, \fthree, \lfd and \lfs, are developed in \mt for efficient evaluations using all of these analytic continuations. Techniques for fast numerical evaluation of the finite summation of series are also employed. The time and the accuracy obtained using these techniques are contrasted with the usual `brute force' summation, and we show that these techniques result in significant improvement in both time and accuracy. The packages are built in a way that the evaluation can be carried out for both generic and non-generic cases.  A comparison of the results using our packages with publicly available implementation in \mt and \textsc{Maple} is also attempted, wherever possible. We find that our packages offer better reliability than these implementations.  A variety of numerical checks are also performed to check the consistency of the package for both generic and non-generic cases. These non-generic cases have further physical applications in Feynman integral calculus, which are discussed in detail.

In the present version, the packages are applicable to the real values of their arguments. However, the analytic continuations derived using the method of Olsson are valid for complex-valued arguments as well. We plan to study in this direction in the near future and build programs that can evaluate these functions for complex-valued arguments. Furthermore, other related cousins of Lauricella-Saran $F_{S}^{(3)}$, such as $F_{N}^{(3)}$ which also appears in Feynman integral calculus,  are also worth investigating in the same spirit.



\section*{Acknowledgements}
We thank Prof. B. Ananthanarayan for reading the manuscript carefully and providing valuable suggestions and useful comments. We also thank Sudeepan Datta for technical help with some cross-checks. We also thank Sumit Banik for testing the packages. SB extends sincere gratitude to Prof. Thomas Gehrmann for the hospitality at the University of Zurich, where a part of this work was carried out. SB thanks Prof. Narayan Rana and NISER Bhubaneshwar for their hospitality at the time of completion of this manuscript. S.B. has been supported by an appointment to the JRG Program at the
APCTP through the Science and Technology Promotion Fund and Lottery Fund
of the Korean Government and by the Korean Local Governments –
Gyeongsangbuk-do Province and Pohang City. The Yukawa Postdoctoral Fellowship of TP is supported by the Yukawa Memorial Foundation and JST CREST (Grant No. JPMJCR19T2). TP is also grateful to the long-term workshop YITP-T-23-01 held at YITP, Kyoto University, where a part of this work was done.

\appendix

\section{Definitions of the relevant series} \label{appendix:def}

This section contains the definitions of the other hypergeometric series that appear in the analytic continuations of the functions considered in this paper. 

\begin{align*}
&G_2\left(a, a^{\prime} , b, b^{\prime} \left|x,y\right.\right)=\sum_{m, n=0}^{\infty}(a)_m\left(a^{\prime}\right)_n(b)_{n-m}\left(b^{\prime}\right)_{m-n} \frac{x^m }{m !} \frac{y^n}{n !} \\
& F_{5c}(a_1,a_2,a_3,a_4,b;c\left|x,y,z\right.) =  \sum_{m,n,p = 0}^\infty \frac{\left(a_1\right)_{m+n}\left(a_2\right)_m\left(a_3\right)_p\left(a_4\right)_p(b)_{n-p}}{(c)_{m+n}}   \frac{x^m }{m !} \frac{y^n}{n !}  \frac{z^p}{p!}\\
&F_{1e}(a_1,a_2,a_3, b_1, b_2, b_3\left|x,y,z\right.) =  \sum_{m,n,p = 0}^\infty \left(a_1\right)_m\left(a_2\right)_p\left(a_3\right)_p\left(b_1\right)_{m-n}\left(b_2\right)_{n-p}\left(b_3\right)_{n-m} \frac{x^m }{m !} \frac{y^n}{n !}  \frac{z^p}{p!}\\
& F_{5b}(a_1,a_2,a_3,a_4,b;c\left|x,y,z\right.) =  \sum_{m,n,p = 0}^\infty \frac{\left(a_1\right)_m\left(a_2\right)_n\left(a_3\right)_p\left(a_4\right)_p(b)_{m+n-p}}{(c)_{m+n}} \frac{x^m }{m !} \frac{y^n}{n !}  \frac{z^p}{p!}\\
&F_{4h}(a_1,a_2,a_3,b_1,b_2;c\left|x,y,z\right.)  = \sum_{m,n,p = 0}^\infty \frac{\left(a_1\right)_{m+n}\left(a_2\right)_n\left(a_3\right)_p\left(b_1\right)_{m-p}\left(b_2\right)_{p-m}}{(c)_n}\frac{x^m }{m !} \frac{y^n}{n !}  \frac{z^p}{p!}\\
& F_G(a_1,a_2,a_3,a_4;c_1,c_2\left|x,y,z\right.) =   \sum_{m,n,p = 0}^\infty 
\frac{\left(a_1\right)_{m+n+p}\left(a_2\right)_m\left(a_3\right)_n\left(a_4\right)_p}{\left(c_1\right)_{m+n}\left(c_2\right)_p} \frac{x^m }{m !} \frac{y^n}{n !}  \frac{z^p}{p!}\\
&F_M(a_1,a_2,b_1,b_2;c_1,c_2\left|x,y,z\right.) = \sum_{m,n,p = 0}^\infty \frac{\left(a_1\right)_{m+n}\left(a_2\right)_{n+p}\left(a_3\right)_m\left(a_4\right)_p}{\left(c_1\right)_{m+n}\left(c_2\right)_p} \frac{x^m }{m !} \frac{y^n}{n !}  \frac{z^p}{p!}
\end{align*}

The series $F_G, F_M$ are introduced by Lauricella \cite{LauricellaFs} and Saran \cite{Saran_transhypgeo}. The series $F_{5b}$ is introduced by Exton \cite{Exton:1973b}. In his notation, it is denoted as $D^{1,3}_3$.

We have followed \cite{Srivastava:1985} for these definitions. It may be noted that a permutation of variables may be required to recover the original definitions. 

\section{Error estimation}\label{appendix:error}

As the packages evaluate the value the functions by summing  suitable analytic continuations, the error in the value depends on the upper limit of the summation (\texttt{N}). In this Section, we perform tests regarding the error estimation of the values obtained by the packages.

For each of the four functions, we randomly choose two sets of  Pochhammer parameters and arguments. For a given precision, we evaluate these series by summing them with varying upper limit of summation \texttt{N}. 
We observe that, as \texttt{N} tends to increase, the values of these summations tend to approach the true value for the chosen precision. The \% Error is calculated as
\begin{align*}
    \% \text{Error} = \left| \frac{\text{true value} -  \text{value obtained upper summation limit \texttt{N} }}{\text{true value}}\right| \times 100
\end{align*}

The data are recorded in Table \ref{table:f1error} and \ref{table:f3error} for the double variable Appell $F_1$ and $F_3$ functions respectively. The chosen precision for these tests are taken to be 50. 
For instance, in Table \ref{table:f1error}, we evaluate Appell $F_1(1/3,1/5,1/2;3/7~|5/7,11/5)$ for 50 significant digits with \texttt{N} varying  from 10 to 90. We note that, the true value of the function is already obtained with $\texttt{N}= 80$. Indeed, summing the series with $\texttt{N}=90$, does not alter the value of the function for the given precision. In a similar manner for $F_3$ in Table \ref{table:f3error}, the true value is obtained with $\texttt{N} = 300$. We ensure that, this is indeed the true value by evaluating $F_3$ with $\texttt{N}= 310$, which is not shown explicitly in the Table.

We repeat a similar analysis for the case of three variable Lauricella $F_{D}^{(3)}$ and Lauricella-Saran $F_{S}^{(3)}$ function with 30 significant digits, the results of which are provided in Table \ref{table:fderror1}, \ref{table:fderror2} and \ref{table:fserror1}, \ref{table:fserror2} respectively.  In spite the computational complexity involved in the case of three variables, we observe that the results are precise even for smaller values of \texttt{N}, say 20, and ensure that the results are precise to at least 1 part in $10^{6}$. We observe from the tables that the values converge quickly as is evident from the rapid decrease of the relative error percentage in the third column.

With these tests, we conclude that the availability of a large number of analytic continuations and the summation technique used to perform the summations are very advantageous for the package as it allows for very fast and efficient evaluation of the  functions discussed in the paper even with small value of upper limit of the summation \texttt{N}.

\begin{table}[]
\begin{tabular}{|l|ll|l|}
\hline
\multicolumn{1}{|c|}{Parameters}                                                                   & \multicolumn{1}{c|}{\texttt{N}}          & \multicolumn{1}{c|}{Numerical Values}                                                                                                                                                                             & \multicolumn{1}{c|}{\% Error} \\ \hline

\multirow{11}{*}{\begin{tabular}[c]{@{}l@{}} \\\\$a=\frac{1}{3}$,\\\\ 
$b_1 = \frac{1}{5}$,\\ \\
$b_2=\frac{1}{2}$,\\ \\
$c=\frac{3}{7}$,\\ \\
$x=\frac{5}{7}$,\\\\
$y=\frac{11}{5}$\end{tabular}} & \multicolumn{1}{c|}{10}  & \begin{tabular}[c]{@{}l@{}}0.32237963682376698631803494051284977821005839050644\\ -1.0582059021814208010155875189053131675953790939882 $i$\end{tabular}  & $2.04 \times 10^{-8}$         \\ \cline{2-4}                             & \multicolumn{1}{c|}{20}   & \begin{tabular}[c]{@{}l@{}}0.32237963665843734682281309218340085706662395739927\\ -1.0582059020280012508949999013616153299520008301095 $i$\end{tabular}  & $6.74 \times 10^{-15}$        \\ \cline{2-4}                             & \multicolumn{1}{c|}{30}   & \begin{tabular}[c]{@{}l@{}}0.32237963665843730512682670889591210682728157063315\\ -1.0582059020280011891345142423183069516332810196575 $i$\end{tabular}  & $5.84 \times 10^{-21}$        \\ \cline{2-4}                              & \multicolumn{1}{c|}{40}   & \begin{tabular}[c]{@{}l@{}}0.32237963665843730512681200642071175445463111522772\\ -1.0582059020280011891344513863908446834887037800833 $i$\end{tabular}  & $1.00 \times10^{-26}$         \\ \cline{2-4}                             & \multicolumn{1}{c|}{50}   & \begin{tabular}[c]{@{}l@{}}0.32237963665843730512681200641472950324320330846929\\ -1.0582059020280011891344513862800100791916244597606 $i$\end{tabular}  & $2.24 \times 10^{-32}$        \\ \cline{2-4}                              & \multicolumn{1}{c|}{ 60}  & \begin{tabular}[c]{@{}l@{}}0.32237963665843730512681200641472950060667875285736\\ -1.0582059020280011891344513862800098318311957874644 $i$\end{tabular}  & $5.55 \times 10^{-38}$        \\ \cline{2-4}                              & \multicolumn{1}{c|}{70}   & \begin{tabular}[c]{@{}l@{}}0.32237963665843730512681200641472950060667753001702\\ -1.0582059020280011891344513862800098318305817672066 $i$\end{tabular}  & $1.47 \times 10^{-43}$        \\ \cline{2-4}                             & \multicolumn{1}{c|}{80}  & \begin{tabular}[c]{@{}l@{}}0.32237963665843730512681200641472950060667753001644\\ -1.0582059020280011891344513862800098318305817655853 $i$\end{tabular}  & 0                             \\ \cline{2-4}                              & \multicolumn{1}{c|}{ 90}  & \begin{tabular}[c]{@{}l@{}}0.32237963665843730512681200641472950060667753001644\\ -1.0582059020280011891344513862800098318305817655853 $i$\end{tabular}  & 0                             \\ 
\cline{1-4}

\multirow{10}{*}{\begin{tabular}[c]{@{}l@{}}\\\\
$a=\frac{1}{2}$,\\\\
$b_1 = \frac{1}{5}$,\\\\
$b_2 = \frac{2}{3}$,\\\\
$c = 2$,\\\\
$x = -\frac{15}{7}$,\\\\
$y=\frac{131}{29}$
\end{tabular}}                                               & \multicolumn{1}{c|}{ 10}  & \begin{tabular}[c]{@{}l@{}}0.73617716270929725507707191392365203376666638544055\\ -0.66270326012045220074083425912188241654098812384932 $i$\end{tabular} & $0.000337$                    \\ \cline{2-4}                              & \multicolumn{1}{c|}{20}   & \begin{tabular}[c]{@{}l@{}}0.73617715411010566658246122196964001353065840347352\\ -0.66269992574422879991084448705094648890438100960316 $i$\end{tabular} & $9.27 \times 10^{-8}$         \\ \cline{2-4}                             & \multicolumn{1}{c|}{30}   & \begin{tabular}[c]{@{}l@{}}0.73617715410923660070804941823231992682655174170477\\ -0.66269992482655115156205693752697891181901895246268 $i$\end{tabular} & $3.43 \times 10^{-11}$        \\ \cline{2-4}                              & \multicolumn{1}{c|}{40}   & \begin{tabular}[c]{@{}l@{}}0.73617715410923643184259278513319864125934297160350\\ -0.66269992482621166461508288326476162220185536233043 $i$\end{tabular} & $1.44 \times 10^{-14}$        \\ \cline{2-4}                             & \multicolumn{1}{c|}{50}   & \begin{tabular}[c]{@{}l@{}}0.73617715410923643179975877678971852483225650798289\\ -0.66269992482621152216674653427730922467533071281496 $i$\end{tabular} & $6.48 \times 10^{-18}$        \\ \cline{2-4}                             & \multicolumn{1}{c|}{ 60}  & \begin{tabular}[c]{@{}l@{}}0.73617715410923643179974619258853491690223665782427\\ -0.66269992482621152210263232044121166238382486541333$i$\end{tabular}  & $3.05 \times 10^{-21}$        \\ \cline{2-4} 
& \multicolumn{1}{c|}{70}   & \begin{tabular}[c]{@{}l@{}}0.73617715410923643179974618853041651715924174080820\\ -0.66269992482621152210260214214077266067556552830793$i$\end{tabular}  & $1.48 \times 10^{-24}$        \\ \cline{2-4}                              & \multicolumn{1}{c|}{80}  & \begin{tabular}[c]{@{}l@{}}0.73617715410923643179974618852902081419773725603447\\ -0.66269992482621152210260212748760933764760831151237$i$\end{tabular}  & $7.35 \times 10^{-28}$        \\ \cline{2-4} 
& \multicolumn{1}{c|}{ 90}  & \begin{tabular}[c]{@{}l@{}}0.73617715410923643179974618852902031098060792005555\\ -0.66269992482621152210260212748033008763832081115411$i$\end{tabular}  & $3.72 \times 10^{-31}$        \\ \cline{2-4}                              & \multicolumn{1}{c|}{100} & \begin{tabular}[c]{@{}l@{}}0.73617715410923643179974618852902031079251058001085\\ -0.66269992482621152210260212748032640770232080358386$i$\end{tabular}  & $1.91 \times 10^{-34}$        \\  \cline{2-4}                               & \multicolumn{1}{c|}{200} & \begin{tabular}[c]{@{}l@{}}0.73617715410923643179974618852902031079243821235572\\ -0.66269992482621152210260212748032640581508268352454 $i$\end{tabular} & 0                             \\ \hline
\end{tabular}
\caption{Table for error analysis of  $F_{1}$}\label{table:f1error}
\end{table}

\begin{table}[]
\begin{tabular}{|l|ll|l|}
\hline
Parameters                                                    & \multicolumn{1}{c|}{\texttt{N}}
& \multicolumn{1}{c|}{Numerical Values}
& \multicolumn{1}{c|}{\% Error} \\ \hline
\multirow{11}{*}{\begin{tabular}[c]{@{}l@{}}
\\
\\
$a_1 = \frac{1}{3}$,\\ \\
$a_2 = \frac{1}{5}$,\\ \\

$b_1 = \frac{1}{7}$,\\
\\
$b_2 = \frac{11}{13}$,\\
\\
$c=\frac{3}{2}$,\\\\ 

$x = \frac{17}{11}$,\\
\\
$ y =\frac{97}{23}$\end{tabular}} & \multicolumn{1}{c|}{10}  & \begin{tabular}[c]{@{}l@{}}0.93202784284474114638161794800492356370918578711269\\ -0.54959821023279401659224226446687686745501126760712 $i$\end{tabular} & 0.0164                        \\ \cline{2-4}                         & \multicolumn{1}{c|}{20}   & \begin{tabular}[c]{@{}l@{}}0.93189563680697511629135826542217754641988346814245\\ -0.54948074180198554628136391380262581195775776534051$i$\end{tabular}  & 0.0000155                     \\ \cline{2-4}                
& \multicolumn{1}{c|}{30}   & \begin{tabular}[c]{@{}l@{}}0.93189560133611005792355886849367971042301172994619\\ -0.54948057786998422405535546901305319373620368366614$i$\end{tabular}  & $1.45 \times 10^{-7}$         \\ \cline{2-4}                         & \multicolumn{1}{c|}{40}   & \begin{tabular}[c]{@{}l@{}}0.93189560287154762621099071688766846375457098285872\\ -0.54948057771323774980431447776347754665045404938937$i$\end{tabular}  & $1.97 \times 10^{-9}$        \\ \cline{2-4}                         & \multicolumn{1}{c|}{50}   & \begin{tabular}[c]{@{}l@{}}0.93189560289257318423563796804174075140811108633844\\ -0.54948057771444297295251473292338130059797941231498$i$\end{tabular}  & $2.52 \times 10^{-11}$        \\ \cline{2-4}                         & \multicolumn{1}{c|}{60}  & \begin{tabular}[c]{@{}l@{}}0.93189560289284116415473773342656898263016020556025\\ -0.54948057771446388510207920968837432243772241198495$i$\end{tabular}  & $3.36 \times 10^{-13}$        \\ \cline{2-4}                         & \multicolumn{1}{c|}{70}   & \begin{tabular}[c]{@{}l@{}}0.93189560289284473636459244891452424681664915118146\\ -0.54948057771446417231567106172735934225539468861723$i$\end{tabular}  & $4.70 \times 10^{-15}$        \\ \cline{2-4}                         & \multicolumn{1}{c|}{ 80}  & \begin{tabular}[c]{@{}l@{}}0.93189560289284478627223250590507836072363689197026\\ -0.54948057771446417631291079213722642612095531197558$i$\end{tabular}  & $6.81 \times 10^{-17}$        \\ \cline{2-4}                         & \multicolumn{1}{c|}{ 90}  & \begin{tabular}[c]{@{}l@{}}0.93189560289284478627223250590507836072363689197026\\ -0.54948057771446417631291079213722642612095531197558$i$\end{tabular}  & $6.81 \times 10^{-17}$        \\ \cline{2-4}                         & \multicolumn{1}{c|}{ 100} & \begin{tabular}[c]{@{}l@{}}0.93189560289284478700691255573841074181427171500130\\ -0.54948057771446417637132920577498962705166047532828$i$\end{tabular}  & $1.56 \times 10^{-20}$        \\ \cline{2-4}                         & \multicolumn{1}{c|}{ 300} & \begin{tabular}[c]{@{}l@{}}0.93189560289284478700708043974406900406703816000709\\ -0.54948057771446417637134240274717736932453975785447 $i$\end{tabular} & 0                             \\ \hline
\multirow{10}{*}{\begin{tabular}[c]{@{}l@{}}
\\ \\

$a_1 = 2$,\\
\\
$a_2 = -\frac{1}{2}$,\\
\\
$b_1 = \frac{1}{5}$,\\
\\
$b_2 = \gamma_{E} $,\\
\\
$c = e$,\\\\ 

$x = \frac{4}{3}$,\\
\\
$y = -\frac{29}{11}$\end{tabular}}                       & \multicolumn{1}{c|}{ 10}  & \begin{tabular}[c]{@{}l@{}}1.5962000427627541522893411710737970740499593476497\\ -0.54383777823887444630368789017560233610957435564983$i$\end{tabular}   & 0.00564                       \\ \cline{2-4}                         & \multicolumn{1}{c|}{20}   & \begin{tabular}[c]{@{}l@{}}1.5961049883477590407505764783180101186893004108533\\ -0.54383908236150545626904263666616921952385079610952$i$\end{tabular}   & $2.37 \times 10^{-7}$         \\ \cline{2-4}                         & \multicolumn{1}{c|}{30}   & \begin{tabular}[c]{@{}l@{}}1.5961049884299449715563681078762100685217915226399\\ -0.54383908634084649925553077948403965176265591266084$i$\end{tabular}   & $1.40 \times 10^{-9}$         \\ \cline{2-4}                         & \multicolumn{1}{c|}{40}   & \begin{tabular}[c]{@{}l@{}}1.5961049884312887635668716798076057598394274373386\\ -0.54383908636422116459546344051343930616627542542243$i$\end{tabular}   & $1.08 \times 10^{-11}$        \\ \cline{2-4}                         & \multicolumn{1}{c|}{50}   & \begin{tabular}[c]{@{}l@{}}1.5961049884312991704830499199502763663681920874228\\ -0.54383908636440216296546984902955028655484348238581$i$\end{tabular}   & $9.80 \times 10^{-14}$        \\ \cline{2-4}                         & \multicolumn{1}{c|}{ 60}  & \begin{tabular}[c]{@{}l@{}}1.5961049884312992644121513281204855347438882079529\\ -0.54383908636440379659195533740075090014494101176173$i$\end{tabular}   & $9.76 \times 10^{-16}$        \\ \cline{2-4}                         & \multicolumn{1}{c|}{70}   & \begin{tabular}[c]{@{}l@{}}1.5961049884312992653468019103299264804790146417346\\ -0.54383908636440381284751323036152189033360574084356$i$\end{tabular}   & $1.04 \times 10^{-17}$        \\ \cline{2-4}                         & \multicolumn{1}{c|}{80}  & \begin{tabular}[c]{@{}l@{}}1.5961049884312992653567527324902245640700903790376\\ -0.54383908636440381302057915337941035606967564478672$i$\end{tabular}   & $1.16 \times 10^{-19}$        \\ \cline{2-4}                         & \multicolumn{1}{c|}{ 90}  & \begin{tabular}[c]{@{}l@{}}1.5961049884312992653567527324902245640700903790376\\ -0.54383908636440381302057915337941035606967564478672$i$\end{tabular}   & $1.16 \times10^{-19}$         \\ \cline{2-4}                         & \multicolumn{1}{c|}{100} & \begin{tabular}[c]{@{}l@{}}1.5961049884312992653568653581283126840827407777868\\ -0.54383908636440381302253795233061332479943333986719 $i$\end{tabular}  & $1.62 \times 10^{-23}$        \\ \hline                              & \multicolumn{1}{c|}{300} & \begin{tabular}[c]{@{}l@{}}1.5961049884312992653568653738145914025108598762101\\ -0.54383908636440381302253822514830229446147356322246 $i$\end{tabular}  & 0                             \\ \hline
\end{tabular}
\caption{Table for error analysis of  $F_{3}$}\label{table:f3error}
\end{table}

\begin{table}[]
\centering
\begin{tabular}{|l|c|c|c|}
\hline
\multicolumn{1}{|c|}{Parameters} & \texttt{N}   & Value                                                                                                           & Error                          \\ \hline
                                 & 10  & \begin{tabular}[c]{@{}c@{}}-11.6863315118001560311492651158\\ +0.0594872758179424201563252013552 $i$\end{tabular} & 0.00028371                     \\ \cline{2-4} 
                                 $a=-7/9$                               & 20  & \begin{tabular}[c]{@{}c@{}}-11.6863645640289705266785495177\\ +0.0594886714040987019947737950812 $i$\end{tabular} & $6.3501 \times 10^{-7}$  \\ \cline{2-4} 
$b_1=-38/13$                              & 30  & \begin{tabular}[c]{@{}c@{}}-11.6863646379713845772269189013\\ +0.0594886736019461018649103738357 $i$\end{tabular} & $2.0160\times 10^{-9}$  \\ \cline{2-4} 
$b_2=5/16$                              & 40  & \begin{tabular}[c]{@{}c@{}}-11.6863646382061045077882748789\\ +0.0594886736079712888929307322432 $i$\end{tabular} & $6.8427\times 10^{-12}$ \\ \cline{2-4} 
$b_3= 8/5$                             & 50  & \begin{tabular}[c]{@{}c@{}}-11.6863646382069007289727044938\\ +0.0594886736079928114890947591356 $i$\end{tabular} & $2.7095\times10^{-14}$ \\ \cline{2-4} 
$c=-4/11$                               & 60  & \begin{tabular}[c]{@{}c@{}}-11.6863646382069038800155673377\\ +0.0594886736079929014000263960160 $i$\end{tabular} & $1.2069\times10^{-16}$ \\ \cline{2-4} 
$x= 13/5$                              & 70  & \begin{tabular}[c]{@{}c@{}}-11.6863646382069038940458166479\\ +0.0594886736079929018167173137567 $i$\end{tabular} & $5.8487\times10^{-19}$ \\ \cline{2-4} 
$y=-10/13$                               & 80  & \begin{tabular}[c]{@{}c@{}}-11.6863646382069038941137831854\\ +0.0594886736079929018187967649313 $i$\end{tabular} & $3.0164\times10^{-21}$ \\ \cline{2-4} 
$z=-51/25$                               & 90  & \begin{tabular}[c]{@{}c@{}}-11.6863646382069038941141336230\\ +0.0594886736079929018188077338800 $i$\end{tabular} & $1.6320\times10^{-23}$ \\ \cline{2-4} 
                                 & 100 & \begin{tabular}[c]{@{}c@{}}-11.6863646382069038941141355186\\ +0.0594886736079929018188077942901 $i$\end{tabular} & $9.2\times10^{-26}$    \\ \cline{2-4} 
                                 & 110 & \begin{tabular}[c]{@{}c@{}}-11.6863646382069038941141355292\\ +0.0594886736079929018188077946345 $i$\end{tabular} & $0.\times10^{-28}$     \\ \cline{2-4} 
                                 & 120 & \begin{tabular}[c]{@{}c@{}}-11.6863646382069038941141355293\\ +0.0594886736079929018188077946365 $i$\end{tabular} & $0.\times10^{-28}$     \\ \cline{2-4} 
                                 & 130 & \begin{tabular}[c]{@{}c@{}}-11.6863646382069038941141355293\\ +0.0594886736079929018188077946365 $i$\end{tabular} & $0.\times10^{-28}$    \\ \hline
\end{tabular}
\caption{Table for error analysis of $F_D^{(3)}$} \label{table:fderror1}
\end{table}

\begin{table}[]
\centering
\begin{tabular}{|c|c|c|}
\hline
\texttt{N}  & Values                                                                                                        & Error                          \\ \hline
10 & \begin{tabular}[c]{@{}c@{}}-123.720358749032387499274041509\\ +56.1381472453581189046124627524 $i$\end{tabular} & $0.0091152$                      \\ \hline
20 & \begin{tabular}[c]{@{}c@{}}-123.714775202875491921872976485\\ +56.1492009388380635962867634111 $i$\end{tabular} & $5.7703\times10^{-8}$  \\ \hline
30 & \begin{tabular}[c]{@{}c@{}}-123.714775162662556753881925146\\ +56.1492010061334435843768863302 $i$\end{tabular} & $1.7619\times10^{-13}$ \\ \hline
40 & \begin{tabular}[c]{@{}c@{}}-123.714775162662429083066395414\\ +56.1492010061336460607279148135 $i$\end{tabular} & $3.9011\times10^{-19}$ \\ \hline
50 & \begin{tabular}[c]{@{}c@{}}-123.714775162662429082778341298\\ +56.1492010061336460611728019397 $i$\end{tabular} & $7.20\times10^{-25}$   \\ \hline
60 & \begin{tabular}[c]{@{}c@{}}-123.714775162662429082778340760\\ +56.1492010061336460611728027572 $i$\end{tabular} & $0.\times10^{-28}$     \\ \hline
70 & \begin{tabular}[c]{@{}c@{}}-123.714775162662429082778340760\\ +56.1492010061336460611728027572 $i$\end{tabular} & $0.\times10^{-28}$     \\ \hline
\end{tabular}
\caption{Table for error analysis of $F_D^{(3)}\left(-\frac{15}{8},\frac{11}{5},\frac{32}{13},-\frac{45}{23};-\frac{3}{7}|\frac{4}{3},-\frac{1}{12},-\frac{17}{7}\right)$ }  \label{table:fderror2}
\end{table}

\begin{table}[]
\centering
\begin{tabular}{|c|c|c|}
\hline
\texttt{N} & Values                                                                                                          & Errors                         \\ \hline
10         & \begin{tabular}[c]{@{}c@{}}-0.803908163357873217155382056555\\ +0.211003425885467902353282840016 $i$\end{tabular} & $0.00021471$                     \\ \hline
20         & \begin{tabular}[c]{@{}c@{}}-0.803906383268229614511276388884\\ +0.211003435589631955611147021355 $i$\end{tabular} & $5.3436\times10^{-7}$  \\ \hline
30         & \begin{tabular}[c]{@{}c@{}}-0.803906378862638363164442787431\\ +0.211003435590046775565175703797 $i$\end{tabular} & $4.2892\times10^{-9}$  \\ \hline
40         & \begin{tabular}[c]{@{}c@{}}-0.803906378827447548532049338492\\ +0.211003435590046871020506360545 $i$\end{tabular} & $5.5092\times10^{-11}$ \\ \hline
50         & \begin{tabular}[c]{@{}c@{}}-0.803906378826997267052725587482\\ +0.211003435590046871053896046887 $i$\end{tabular} & $9.1573\times10^{-13}$ \\ \hline
60         & \begin{tabular}[c]{@{}c@{}}-0.803906378826989805067057363331\\ +0.211003435590046871053911227118 $i$\end{tabular} & $1.7929\times10^{-14}$ \\ \hline
70         & \begin{tabular}[c]{@{}c@{}}-0.803906378826989659317483249137\\ +0.211003435590046871053911235198 $i$\end{tabular} & $3.9325\times10^{-16}$ \\ \hline
80         & \begin{tabular}[c]{@{}c@{}}-0.803906378826989656126996327034\\ +0.211003435590046871053911235202 $i$\end{tabular} & $9.3756\times10^{-18}$ \\ \hline
90         & \begin{tabular}[c]{@{}c@{}}-0.803906378826989656051053102167\\ +0.211003435590046871053911235202 $i$\end{tabular} & $2.3829\times10^{-19}$ \\ \hline
100        & \begin{tabular}[c]{@{}c@{}}-0.803906378826989656049125531115\\ +0.211003435590046871053911235202 $i$\end{tabular} & $6.3708\times10^{-21}$ \\ \hline
110        & \begin{tabular}[c]{@{}c@{}}-0.803906378826989656049074055667\\ +0.211003435590046871053911235202 $i$\end{tabular} & $1.7747\times10^{-22}$ \\ \hline
120        & \begin{tabular}[c]{@{}c@{}}-0.803906378826989656049072623175\\ +0.211003435590046871053911235202 $i$\end{tabular} & $5.114\times10^{-24}$  \\ \hline
130        & \begin{tabular}[c]{@{}c@{}}-0.803906378826989656049072581927\\ +0.211003435590046871053911235202 $i$\end{tabular} & $1.52\times10^{-25}$   \\ \hline
140        & \begin{tabular}[c]{@{}c@{}}-0.803906378826989656049072580705\\ +0.211003435590046871053911235202 $i$\end{tabular} & $5.\times10^{-27}$     \\ \hline
150        & \begin{tabular}[c]{@{}c@{}}-0.803906378826989656049072580667\\ +0.211003435590046871053911235202 $i$\end{tabular} & $0.\times10^{-28}$     \\ \hline
160        & \begin{tabular}[c]{@{}c@{}}-0.803906378826989656049072580666\\ +0.211003435590046871053911235202 $i$\end{tabular} & $0.\times10^{-28}$     \\ \hline
\end{tabular}
\caption{Table for error analysis of $F_S^{(3)}\left(-\frac{19}{8},\frac{10}{7},-\frac{4}{11},-\frac{5}{3},\frac{1}{11};\frac{1}{2}|\frac{8}{11},\frac{11}{14},\frac{24}{11}\right)$  }\label{table:fserror1}
\end{table}

\begin{table}[]
\centering
\begin{tabular}{|c|c|c|}
\hline
\texttt{N}  & Values                                                                                                       & Error                          \\ \hline
10 & \begin{tabular}[c]{@{}c@{}}9.16305265536863465417794986565\\ +8.27584751678110553098267576480 $i$\end{tabular} & $0.000017284$                    \\ \hline
20 & \begin{tabular}[c]{@{}c@{}}9.16305102898786458345532292326\\ +8.27584613541487342678297690304 $i$\end{tabular} & $1.6538\times10^{-9}$  \\ \hline
30 & \begin{tabular}[c]{@{}c@{}}9.16305102882763612514913095123\\ +8.27584613528828931677136687240 $i$\end{tabular} & $5.9850\times10^{-14}$ \\ \hline
40 & \begin{tabular}[c]{@{}c@{}}9.16305102882762933435112049963\\ +8.27584613528829222943252462630 $i$\end{tabular} & $6.1805\times10^{-17}$ \\ \hline
50 & \begin{tabular}[c]{@{}c@{}}9.16305102882762933672160252724\\ +8.27584613528829223668060615117 $i$\end{tabular} & $4.2635\times10^{-20}$ \\ \hline
60 & \begin{tabular}[c]{@{}c@{}}9.16305102882762933672331780071\\ +8.27584613528829223668557904573 $i$\end{tabular} & $3.0433\times10^{-23}$ \\ \hline
70 & \begin{tabular}[c]{@{}c@{}}9.16305102882762933672331885613\\ +8.27584613528829223668558264889 $i$\end{tabular} & $2.5\times10^{-26}$    \\ \hline
80 & \begin{tabular}[c]{@{}c@{}}9.16305102882762933672331885687\\ +8.27584613528829223668558265184 $i$\end{tabular} & $0.\times10^{-28}$     \\ \hline
90 & \begin{tabular}[c]{@{}c@{}}9.16305102882762933672331885687\\ +8.27584613528829223668558265184 $i$\end{tabular} & $0.\times10^{-28}$     \\ \hline
\end{tabular}

\caption{Table for error analysis of $F_S^{(3)} \left( -\frac{23}{10},-\frac{5}{4},-\frac{10}{7},\frac{24}{25},\frac{3}{5};-\frac{18}{7}|-\frac{5}{6},\frac{29}{12},\frac{27}{10}\right)$}\label{table:fserror2}
\end{table}
\newpage

\bibliography{references.bib}

\end{document}